\documentclass[published]{JHEP}

\usepackage{epsfig}
\usepackage{amssymb}
\usepackage{bbm}

\newcommand{\sign}{\mathop{\rm sign}\nolimits}
\newcommand{\dett}{\mathop{\rm det}\nolimits}
\newcommand{\Ree}{\mathop{\rm Re}\nolimits}
\newcommand{\identity}{\mathbbm{1}}
\newcommand{\dbar}{\partial \hskip -5.5pt/ }

\title{Preheating of massive fermions after inflation:
analytical results}

\author{Marco Peloso and Lorenzo Sorbo\\
SISSA/ISAS, Trieste, Italy, via Beirut 2-4, 34013\\
INFN Trieste, Italy, via Valerio 2, 34127\\
E-mail: \email{peloso@sissa.it}, \email{sorbo@sissa.it}}

\abstract{Non-perturbative production of fermions after chaotic inflation has
been the object of several studies in the very recent past.
However, the results in the most interesting case of production of
massive fermions in an expanding Universe were so far known only
numerically. We provide very simple and readable analytical
formulae, both for the spectra of the created fermions and for
their total energy density. Their derivation is closely related to
the one adopted for bosons and exploits the fact that the
production occurs during very short intervals of non-adiabatical
change of the fermionic frequency. Our formulae show the presence
of resonance bands if the expansion of the Universe is neglected,
and their disappearance when the latter is included. As in the
bosonic case, this last effect is due to the stochastic character
that the expansion gives to the whole process. Backreaction is
considered in the last part of the work. All our analytical
results are in excellent agreement with the previous numerical
ones in the regime of validity of the latter. However, a more
accurate scaling for the energy density of the produced fermions
is here found.}


\received{March 7, 2000}

\accepted{April 6, 2000}

\JHEP{05(2000)016}

\begin{document}

\section{Introduction}\label{sec:1}

One of the key ingredients for our understanding of the early
Universe is the mechanism of inflation~\cite{guth}, which
constitutes a very elegant solution to several cosmological
problems.

Despite of the simplicity of the general idea, the details of the
physical processes which govern it are still somehow unclear and
matter of intense work. The two main aims of these studies are
\begin{list}{}{\setlength{\labelwidth}{13pt}}
\item[(i)] to embed inflation in a context more motivated by
particle physics (for a review see~\cite{lr}) and
\item[(ii)] to understand the reheating phase which converts the
energy density that drives inflation into the matter and radiation
that we presently see.
\end{list}

This second issue has been deeply influenced in the last decade by
the possibility of particle creation through parametric
resonance~\cite{tb}. The application of this phenomenon to
creation of matter after inflation has been called
\emph{preheating} in the paper~\cite{kls}, since (with the
exception of some very recent versions~\cite{fkl}) it is usually
followed by a stage of (ordinary) perturbative reheating.

Preheating of bosons~\cite{kls,stb} is characterized by a very
efficient and explosive creation, due to the coherent effect of
the oscillations of the inflaton field. This allows significant
production even when single particle decay is kinematically
forbidden. It has been very recently noticed~\cite{gprt} that
preheating of fermions can also be very efficient despite the
production is in this case limited by Pauli blocking.

Parametric creation of spin $1/2$ fermions has been the subject of
some works in the past. Pure gravitational production has been
examined in refs.~\cite{mmf, kt}, while creation by an oscillating
background field is instead considered in the
works~\cite{gmm,bhp,gk,fkl,gprt,bp}. References~\cite{gmm,bhp} report
results for creation in a Minkowski space. Reference~\cite{gk} studies
the production of massless fermions after a $\lambda  \phi^4$
inflation, exploiting the fact that this case can also be
reconducted to a static one. In this work, production in a static
Universe after chaotic inflation is also considered, and some
conjectures on the effects of the expansion are made. Moreover,
the full calculation of preheating of massive fermions after
chaotic inflation in an expanding Universe has been performed
numerically in ref.~\cite{gprt}.

\looseness=1
These last works had a great impact on the most recent studies.
For example, their results turned also useful to the study of
gravitinos production at preheating~\cite{gra}. This issue is
particularly important, since gravitinos can easily overclose the
Universe (if they are stable) or (if they decay) spoil the
successful predictions of primordial nucleosynthesis through
photodissociation of the light elements. Gravitinos can be
thermally produced during the stage of reheating. To avoid this
overproduction, the reheating temperature $T_{\mathrm{RH}}$ after inflation
cannot be larger than $\sim (10^{8}-10^{9})$~\,~GeV~\cite{grath}.
However, it has been realized~\cite{gra} that the non-thermal
production of helicity $\pm 1/2$ gravitinos (whose equation of
motion can be reconducted to the one of an ordinary spin $1/2$
Dirac particle) can easily be more efficient than the thermal one,
and this in general leads to more stringent upper bounds on
$T_{\mathrm{RH}}$. Several papers related to the works~\cite{gra} have
recently appeared~\cite{gra2,grahyb}.

Another important implication of preheating of fermions is
constituted by leptogenesis, as the work~\cite{gprt} and the
related papers~\cite{lepto} show. In this scheme~\cite{fy}, a
leptonic asymmetry is first created from the decay of
right-handed neutrinos, and then partially converted to baryon
asymmetry through sphaleronic interactions. Since leptogenesis is
very sensitive to both the mechanism of creation of the heavy
neutrinos and to the neutrino mass matrices, it could constitute
an interesting link between preheating and the experiments on
neutrino oscillations.

Other phenomenological implications of these works appear in
refs.~\cite {high}, with preheating as a possible mechanism for creating
superheavy relic particles responsible for the ultrahigh energy
cosmic rays, and in ref.~\cite{ckrt}, where the possible impact of
fermions produced \emph{during} inflation on the microwave
background anisotropies and on the large-structure surveys is
considered. Finally, fermionic production can play an interesting
role in hybrid inflationary models~\cite{grahyb, hyb2}.

Due to the large number of these studies, it may be worth to
reconsider the basic mechanism of fermionic preheating. In
particular, it should be important to give an analytical
confirmation to the results of production of massive fermions in
the expanding Universe, which are so far known only numerically
from the analysis~\cite{gprt}. This is the aim of the present
work.

In the next section we revise the basic formalism for preheating
of fermions. We consider creation of very massive particles right
after chaotic inflation. The coupling of the fermions to the
oscillating inflaton gives them a time varying mass. As it is
known, this can cause a non-adiabatical change of the frequency of
the fermions and their consequent creation. In case of very
massive fermions, the non-adiabaticity condition can be satisfied
only when their total mass vanishes, and the production occurs at
discrete intervals, until the inflaton oscillations become too
small for the total fermionic mass to vanish.

\looseness=-1
In section~\ref{sec:3} we derive analytical formulae for the
spectra of the fermions after a generic production. Our derivation
follows the one developed in ref.~\cite{kls} for preheating of
bosons. It exploits the fact that the production occurs in very
short intervals around the zeros of the total fermionic mass: the
calculation is made possible from the fact that the occupation
number can be considered as constant outside these small regions,
and that the expansion of the Universe can be neglected inside
them. As a result, the only physical quantities relevant for the
creation are the time derivative~$\phi'$ of the inflaton field and
the value of the scale factor $a$ at each production. The
derivation of the ``fermionic counterpart'' of the formulae
obtained for preheating of bosons in ref.~\cite{kls} has also been
done to a certain extent in the work~\cite{ckrt}, where the
results of a single production during inflation is given. However,
when one is interested in the successive productions, a more
detailed study is necessary, as our analysis shows. As may be
expected, the final results that we obtain closely resemble the
ones of the work~\cite{kls}. What is most surprising is their
excellent agreement with the numerical results, as some figures
provided  manifest.

In section~\ref{sec:4} we consider the production in a
non-expanding Universe. In this case our analytical formulae
considerably simplify and agree with the ones of
refs.~\cite{bhp,gk}. In particular, they show the presence of
resonance bands which are anyhow limited by Pauli blocking.

\looseness=1
In section~\ref{sec:5} we study the more interesting case of production in
an expanding Universe. As indicated in ref.~\cite{gk} and as
confirmed by the numerical results of the work~\cite{gprt}, the
creation is now very different with respect to the previous case.
The expansion removes the resonance bands and the production
(almost) saturates a Fermi sphere up to a maximal momentum. Our
analytical results confirm this behavior.

In section~\ref{sec:5} we also calculate the total energy density
$\rho_X$ of produced fermions, which may be the quantity of most
physical relevance. We compare our results with the ones of
ref.~\cite{gprt}, where it is shown that the final value of
$\rho_X$ (normalized to the inflaton energy density) scales
linearly with the parameter $q \equiv( g^2 \phi_0^2) / ( 4
m_\phi^2)$,\footnote{In this expression $g$ is the Yukawa coupling
between the inflaton and the fermions, $\phi_0$ the initial value
of the inflaton, and $m_\phi$ its mass.} while it depends very
weakly on the fermion bare mass $m_X$. However, these numerical
results are valid only in a limited range for $m_X$, and it has
been wondered if the density $\rho_X$ decreases at values of $m_X$
below this range. Our analytical results give a positive answer in
this regard. To see this, a proper average of the analytical
formulae must be done, exploiting the fact that the expansion of
the Universe gives the production a stochastic character. In this
way one can get a ``mean'' function that interpolates very well
between the maxima and the minima of the spectra of produced
particles. Again we derived it in close analogy with what is done
in the bosonic case, and again the results that we get are in very
good agreement with the numerical ones of ref.~\cite{gprt} in the
region of validity of the latter.

All this analysis neglects the backreaction of the produced
fermions on the evolution of the inflaton field and of the scale
factor. Despite the difficulty of a more complete treatment,
backreaction effects can be understood at least in the Hartree
approximation. This was done numerically in ref.~\cite{gprt}. In
section~\ref{sec:6} we see that the analytical formulae here provided
allow to understand the effects of backreaction observed in the
numerical simulations.

\section{Production of fermions at preheating}\label{sec:2}

In this section we revise the basic formalism for production of
fermions at preheating. Our presentation follows the ones of
refs.~\cite{gprt, ckrt} with the correction of few typos.

We start from the Dirac equation (in conformal time $\eta$) for a
fermionic field $X$ in the Friedmann-Robertson-Walker background:
\begin{equation} \label{dir1}
\left( \frac{i}{a}\,\gamma^\mu \,\partial_\mu + i\, \frac{3}{2}
H \gamma^0 - m \right) X = 0\,,
\end{equation}
where $a$ is the scale factor of the Universe, $H=a'/a^2$ the
Hubble rate,\footnote{Here and in the following prime denotes
derivative which respect to $\eta$.} and $m$ the total mass of
the fermion.

Fermionic production is possible if this last quantity varies non
adiabatically with time. This may happen during the coherent
oscillations of the inflaton $\phi$ at reheating in presence of a
Yukawa interaction $\phi {\bar X} X$, such that the
total mass is given by
\begin{equation} \label{mass}
m (\eta)=m_X + g \phi( \eta)\,.
\end{equation}
We will always consider fermions with very high bare mass $m_X$.
In this case, non adiabaticity can be achieved if the Yukawa
interaction is sufficiently strong to make the total
mass~(\ref{mass}) vanish. Fermions are then created whenever the
inflaton field crosses the value $\phi_* \equiv - m_X / g $.

We will study the production after chaotic inflation, that is
while the inflaton field coherently oscillates about the minimum
of the potential
\begin{equation}
V= \frac{1}{2} \,
m_\phi^2  \phi^2\,, \quad m_\phi \simeq 10^{13}\, \mathrm{GeV} \,.
\end{equation}
If one neglects the backreaction of the created particles (this
effect will be considered in the last part of the work), then,
after few oscillations, the inflaton evolves according to the
formula
\begin{equation} \label{infla}
\phi (t) \simeq \frac{M_p}{\sqrt{3\pi}}
\frac{\cos\left( m_\phi  t \right)}{m_\phi  t} \,,
\end{equation}
where $t$ is the physical time. The presence of the $t$ at the denominator
of eq.~(\ref{infla}) shows the damp of the oscillations due to the
expansion of the Universe. It thus follows that there exists a final time
after which $ | \phi | < m_X / g$ and the total mass no longer
vanishes, so that the production ends.

To proceed with our analysis, we redefine $\chi = a^{-3/2}X$.
Equation~(\ref{dir1}) acquires the more familiar form
\begin{equation} \label{dir2}
\left( i \dbar - a  m \right)  \chi
= 0 \,.
\end{equation}

We decompose
\begin{equation} \label{decomp}
\chi ( x ) = \int \frac{d^3 k}{(2\pi )^{3/2}}
e^{-i {\bf k} \cdot \mathbf{x}}
\sum_r \left[ u_r(k, \eta)a_r(k) + v_r(k,\eta)b_r^\dagger
 (-k)\right] ,
\end{equation}
with\footnote{We choose
\begin{eqnarray*}
&\gamma^0 =\pmatrix{\identity & 0 \cr 0 & -\identity},\qquad
\gamma^1 = \pmatrix{ 0 & -\,i\,\sigma_2 \cr -\,i\,\sigma_2 &0 },
\qquad\gamma^2 = \pmatrix{ 0 & i\sigma_1 \cr i \sigma_1 & 0},
\\&
\gamma^3 = \pmatrix{ 0 & \identity \cr -\identity & 0} 
\longrightarrow C = \pmatrix{0 & -\sigma_1 \cr\sigma_1 & 0 }\,.
\end{eqnarray*}}
\begin{equation} \label{conj}
v_r ( k ) = C {\bar u}_r^T \left( -  k \right).
\end{equation}

It is convenient to take the momentum along the third
direction $k \equiv k_z$ and to~define
\begin{equation} \label{spin}
u_r \equiv \left[ \frac{u_+ ( \eta)}
{\sqrt{2}}  \psi_r \,, \frac{u_- (\eta)}{\sqrt{2}} \psi_r \right]^T \,,
\qquad v_r \equiv \left[ \frac{v_+ \left( \eta \right)}
{\sqrt{2}}  \psi_r \,, \frac{v_- \left( \eta \right)}{\sqrt{2}}
\psi_r \right]^T
\end{equation}
with $\psi_{+} =$ {\scriptsize{\smash{$\pmatrix{1 \cr 0}$}}} and $\psi_{-}
=$ {\scriptsize{\smash{$\pmatrix{0 \cr 1}$}}} eigenvectors of the helicity
operator $\mathbf{\sigma} \cdot \mathbf{v} / | { \bf v }|$.
From the normalization adopted in eq.~(\ref{spin}), one can
impose
\begin{eqnarray} \label{norma}
u_r^+  u_s = v_r^+  v_s = \delta_{rs}\,,\qquad
u_r^+  v_s = 0\, ,\qquad
| u_+ |^2 + | u_- |^2 = 2\,.
\end{eqnarray}

With this choice, the Dirac equation~(\ref{dir2}) rewrites
\begin{equation} \label{dir3}
u'_\pm ( \eta) = i k  u_\mp
(\eta) \mp i  a  m  u_\pm (\eta)\,,
\end{equation}
which can be decoupled into\footnote{We do not deal with the
analogous equations for $v_\pm$, since from condition~(\ref{conj})
it simply follows $v_+ (\eta) = - u_-^* (\eta), v_- (\eta) = u_+^*
(\eta)$.}
\begin{eqnarray}
u_\pm'' + \left[ \omega_k^2 \pm i ( a m )' \right]
u_\pm =0 \,, \qquad
\omega^2 ( \eta ) = k^2 + a^2  m^2 \,.
\label{eom}
\end{eqnarray}

Under this decomposition, the hamiltonian of the system
\begin{equation}
H = \frac{1}{a} \int d^3 {\bf x }  \chi^+ (x)\left(-i  \partial_\eta
\right) \chi (x)
\end{equation}
rewrites
\begin{eqnarray}
H &=& \frac{1}{a} \int d^3k
\sum_r \Big\{ E_k \left( \eta \right) \left[
a_r^\dagger \left( k \right) a_r \left( k \right) - b_r \left( k \right)
b_r^\dagger \left( k \right) \right]+
\nonumber \\[-2pt]&&
	\hphantom{\frac{1}{a} \int d^3k\sum_r \Big\{} 
+ F_k \left( \eta \right) b_r \left( - k \right) a_r \left( k \right)
+ F_k^* \left( \eta \right) a_r^\dagger \left( k \right) b_r^\dagger
\left( - \, k \right) \Big\}\,,
\end{eqnarray}
where (using the equations of motion)
\begin{eqnarray}
E_k &=& k \, {\Ree} \left( u_+^*  u_- \right) + a\,  m
\left( 1 - u_+^* u_+ \right), \nonumber\\[3pt]
F_k &=& \frac{k}{2} \left( u_+^2 - u_-^2 \right) + a\,m \,u_+  u_-\,,
\qquad E_k^2 +|F_k|^2 =\omega^2 \,. \label{ef}
\end{eqnarray}

It is always possible to choose an initial configuration with the
hamiltonian in the standard (diagonal) form. If we take, for
example,
\begin{equation}
\label{init}
u_\pm \left( \eta_0 \right) =
\left( 1 \mp \frac{a  m}{\omega} \right)^{1/2}
e^{i  \phi}\,,\quad \phi  \mathrm{\ arbitrary\ phase} \,,
\end{equation}
we have $E_k \big( \eta_0\big) = \omega,$ $F_k \big(\eta_0
\big)=0$.

However, the evolution equations~(\ref{eom}) drive $F_k$ different from
zero and a diagonal form for $H$ can be recovered only after a Bogolyubov
transformation on the creation/annihilation operators
\begin{eqnarray}
{\hat a}( \eta , k ) &\equiv& \alpha_k (\eta ) a ( k ) + \beta_k
(\eta) b^+ ( -k )\,, 
\nonumber\\ 
{\hat b}^+ ( \eta, k ) &\equiv& -
\beta_k^* ( \eta ) a ( k ) + \alpha_k^*( \eta ) b^+ ( -k)\,. 
\label{diag}
\end{eqnarray}

While it is immediate to see that the equal time anticommutation
relations on both sets $\{ a^{(+)}, b^{(+)} \}$ and $\{ {\hat
a}^{(+)}, {\hat b}^{(+)}\}$ enforce
\begin{equation} \label{bog}
| \alpha_k |^2 + | \beta_k |^2 = 1 \,,
\end{equation}
the use of some algebra shows that a diagonalization of the
hamiltonian is possible with the choice
\begin{equation}
\alpha = \beta \, \left( \frac{E+\omega}{F^*} \right),\qquad
| \beta |^2 = \frac{| F |^2}{2 \, \omega
\left( E + \omega \right)} = \frac{\omega - E}{2 \, \omega} \,.
\end{equation}

The time varying  operators ${\hat a}^{(+)}$ and ${\hat b}^{(+)}$
are employed to define time dependent Fock spaces, each of them
built from the zero (quasi)particle state at the time $\eta$:
\begin{equation}
{\hat a}( \eta ) |
0_\eta \rangle = {\hat b} ( \eta )
| 0_\eta \rangle = 0 \,.
\end{equation}

Let us consider the state $| 0_{\eta_0} \rangle \equiv | 0
\rangle$ which describes a system with  zero particles at the initial time
$\eta_0 $. This is no longer true for generic time $\eta >
\eta_0 $, since the occupation number at $\eta$ is defined in
terms of the operators which diagonalize the hamiltonian at that
moment. The particle density per physical volume $V=a^3$ at time
$\eta$ is indeed given by\footnote{Of course the same result is
achieved also for antifermions.}
\begin{eqnarray}
&&n ( \eta ) \equiv \langle 0 | \frac{N}{V} | 0 \rangle =\langle 0 |
\frac{1}{a^3} \sum_{r = \pm 1} \int \frac{d^3 {\bf k}}{\left( 2 \pi
\right)^3}\, {\hat a }_r^+ (\eta, r ) {\hat a }_r ( \eta, r ) | 0
\rangle=
\nonumber\\[2pt]
&&\hphantom{n ( \eta )} =\frac{1}{\pi^2 \, a^3} \int\, dk \, k^2
\left| \beta_k\right|^2 ,
\end{eqnarray}
which is different from zero whenever the hamiltonian is non diagonal
in terms of the initial operators. The occupation number of created
fermions is thus given by $n_k = | \beta_k |^2$ and the
condition~(\ref{bog}) ensures that the Pauli limit $ n_k < 1$ is
respected.

\section{Analytical evaluation of the occupation
number}\label{sec:3}

In this section we calculate analytically the evolution of the
Bogolyubov coefficients~(\ref{diag}) during the oscillations of the
inflaton field after chaotic inflation.

In this derivation, we exploit the fact that, in the regime of very
massive fermions we are interested in, the creation occurs only for
very short intervals about the points $\phi_* \equiv - m_X/g $ where
the total fermionic mass (cfr. eq.~(\ref{mass})) vanishes. As the
perfect agreement with the numerical results will confirm, this
consideration allows one to treat the fermionic production with the
same formalism adopted in the bosonic case~\cite{kls}. While far from
the zeros of the total mass $m$ the Bogolyubov coefficients can be
treated as constant, whenever $\phi$ crosses $\phi_*$ a sudden
variation occurs. Since the interval of production is very narrow, one
can safely neglect the expansion of the Universe during the production
and also linearize the function $\phi(\eta) \simeq \phi_* +\linebreak
\phi' \big( \eta_*\big)\big(\eta - \eta_* \big)$.  As a consequence,
the frequency $\omega$ defined in eq.~(\ref{eom}) acquires the~form
\begin{equation}
\omega^2 \simeq k^2 + a^2 \left( \eta_* \right)  \phi'^2
\left(  \eta_* \right) \; \left( \eta - \eta_* \right)^2,
\end{equation}
and the whole calculation strongly resembles the one for scattering
through a qua\-dra\-tic potential.

The use of this formalism is very well established in case of
production of bosons~\cite{kls}. For what concerns fermions, it has
been recently adopted by Chung et al.~\cite{ckrt} for the study of the
production during inflation. \pagebreak[3] Fermionic production during
inflation is possible only if the coefficient $g$ of the Yukawa
interaction has opposite sign with respect to the inflaton field
during inflation, or the total mass~(\ref{mass}) would never
vanish.\footnote{ Of course this constraint does not apply to our
case, since the inflaton field changes sign after each half
oscillation.} If this is the case, it is possible to choose the value
of $g$ such that the production occurs only once during inflation,
while during reheating $| \phi |$ is always too small for having
creation. Our derivation is strongly inspired by this work. However,
we are interested in couplings for which the production occurs several
times during reheating, and it is not at all guaranteed a priori that
an analytical approximation may work also in this case. The present
analysis not only positively answers to this question, but also
provides very simple and explicit formulae valid for arbitrary number
of productions.

As may be expected, the final formulae are very similar to the
ones found for production of bosons~\cite{kls}. What is most
surprising is the almost perfect agreement between the analytical
results and the numerical simulations, as we will show with some
figures at the end of this section.

The first step for the derivation of the analytical formulae is to
consider asymptotic solutions of eqs.~(\ref{dir3})
and~(\ref{eom}). We look for solutions valid for $\phi$ not very
close to $\phi_*$, where the adiabaticity condition $\omega' \ll
\omega^2$ holds.

In this regime, the most general solutions of eqs.~(\ref{dir3})
and~(\ref{eom}) are
\begin{eqnarray}
u_+ &=& A \left( 1 - \frac{a m}{\omega} \right)^{1/2} e^{i \int^\eta
\omega_k \, d \eta} + B \, \left( 1 + \frac{a m}{\omega} \right)^{1/2}
e^{-i \int^\eta \omega_k \, d \eta}\,,
\nonumber \\[2pt]
u_- &=& - B \left( 1 - \frac{a m}{\omega} \right)^{1/2} e^{-i
\int^\eta \omega_k \, d \eta} + A \left( 1 + \frac{a m}{\omega}
\right)^{1/2} e^{i \int^\eta \omega_k \, d \eta} \,, 
\label{adiab}
\end{eqnarray}
with $| A |^2 + | B |^2 = 1$ following from the condition
$| u_+ |^2 + | u_- |^2 = 2$.

We can always put solutions of eqs.~(\ref{dir3}) and~(\ref{eom})
into the form~(\ref{adiab}), with $A$ and $B$ in general functions
of $\eta$. However, in most of the evolution (whenever the
adiabaticity condition holds) it is a very good approximation to
treat the coefficients $A$ and $B$ as constant.

Substituting the expressions~(\ref{adiab}) into eqs.~(\ref{ef}), one finds
\begin{eqnarray}
F = 2  \omega  A  B \,,\qquad
E = \omega \left( | A |^2 - | B |^2 \right),
\end{eqnarray}
from which it follows
\begin{equation}
| \beta |^2 = \frac{| F |^2}{2  \omega
\left( E + \omega \right)}  = | B |^2 \,.
\end{equation}

One can thus choose (up to an irrelevant global phase)
\begin{equation}
A \equiv \alpha  \longrightarrow  B \equiv \beta^* \,,
\end{equation}
where $\alpha$ and $\beta$ are nothing but the Bogolyubov
coefficients we are interested in.

Notice that the initial choice $A \big( \eta_0 \big) =1$,
$B\big(\eta_0 \big) =0$ corresponds to the initial
condition~(\ref{init}) and to the zero particles state chosen in
the previous section.

As we have said, these coefficients undergo a sudden change
whenever $\phi$ crosses~$\phi_0$ and then they stabilize to new
(almost) constant values. Our aim is to find the values at the end
of the variation in terms of the ones prior to it.

We have not specified the lower limit of the integrals appearing
in eqs.~(\ref{adiab}). For present convenience we choose it to be
the time $\eta_{*1}$ of the first production (that is when
$\phi=\phi_*$ for the first time).

Let us consider the evolution equation~(\ref{eom}) near the point
$\eta_{*1}$. Since for high mass $m_X$ the fermionic
production is limited to a very short interval, one can neglect
the expansion of the Universe during it and write the equation for
$ \phi(\eta)$ in a linearized form. We can thus
write
\begin{equation}
a \, m ( \eta ) \simeq a_{*1} \, g\,  \phi_{*1}'
\left( \eta - \eta_{*\,1} \right),\qquad a_{*1} \equiv a
\left( \eta_{*\,1} \right) ,\qquad \phi_{*1} \equiv \frac{d \, \phi}
{d \, \eta} \Big |_{\eta_{*1}} \,.
\end{equation}

Following the notation of~\cite{ckrt}, we define
\begin{equation}
p \equiv \frac{k}{\sqrt{g  | \phi_{*1}' |  a_{*1}}}\, ,\qquad
\tau = \sqrt{g  | \phi_{*1}' | a_{*1}}  \left( \eta - \eta_{*1}
\right).
\end{equation}

In terms of these new quantities, eqs.~(\ref{eom}) rewrite (dot
denotes derivative with respect to $\tau$)\footnote{For the
production at the moment $\eta_{*n}$, when the total mass
vanishes for the n-th time, eq.~(\ref{eqapp}) must be replaced by
\begin{equation}
\ddot{u}_{\pm} + \left( p^2 \pm i \sign \left( \phi_{*n}' \right) +
\tau^2 \right) u_\pm = 0 \,.
\end{equation}
The effect of this replacement on the final results is reported
below.}
\begin{equation} \label{eqapp}
\ddot{u}_{\pm} + \left( p^2 \mp i + \tau^2 \right) u_\pm = 0 \,.
\end{equation}

The point $\eta_{*1}$ is thus mapped into the origin of $\tau$ and
the region of asymptotic adiabaticity is at large $ |\tau|$.

In the asymptotic solutions~(\ref{adiab}) we can see the behaviors
\begin{eqnarray}
\left( 1 + \frac{a  m}{\omega} \right)^{1/2} &\longrightarrow&
\frac{p}{\sqrt{2} \tau}\,, 
\nonumber \\
\left( 1 - \frac{a m}{\omega} \right)^{1/2}
&\longrightarrow& \sqrt{2}\,,
 \nonumber  \\
e^{\pm  i \int^\eta \omega_k d  \eta} &\longrightarrow&
\left( \frac{2  \tau}{p} \right)^{\pm i  p^2/2}  e^{\pm  i  \tau^2/2}
 e^{\pm  i  p^2/4} \,,
\end{eqnarray}

\pagebreak[3]

\noindent for $\tau \rightarrow  + \infty$, and
\begin{eqnarray}
\left( 1 + \frac{a  m}{\omega} \right)^{1/2}
&\longrightarrow& \sqrt{2}\,, 
\nonumber \\
\left( 1 - \frac{a  m}{\omega} \right)^{1/2}
&\longrightarrow& \frac{p}{-\sqrt{2}\tau}\,, 
\nonumber \\
e^{\pm  i \int^\eta \omega_k d  \eta} &\longrightarrow&
\left( \frac{p}{-  2  \tau} \right)^{\pm i  p^2/2}  e^{\mp  i
\tau^2/2}  e^{\mp  i  p^2/4} \,,
\end{eqnarray}
for $\tau \rightarrow  - \infty$.

Equations~(\ref{eqapp}) are solved~\cite{ckrt} by parabolic
cylinder functions $D_\lambda (z)$~\cite{gr}. More precisely, the
combination that matches the asymptotic solution~(\ref{adiab}) at
$\tau\rightarrow -\infty$~is\footnote{We deal only with the
function $u_+$, since the study of $u_-$ leads to the same
results.}
\begin{eqnarray}
&&u_+ ( \tau ) = A^-
\sqrt{2}  \left( \frac{p}{\sqrt{2}} \right)^{1+i p^2/2}
e^{i  {\smash( \frac{\pi}{4} -\frac{p^2}{4})}}
e^{-\pi  p^2/8}  D_{-1-i p^2/2} \left( -
\left( 1 + i \right) \tau \right) + \nonumber\\[5pt]
&&\hphantom{u_+ ( \tau )=}
+ B^-  \sqrt{2}  \left( \frac{p}{\sqrt{2}} \right)^{- i p^2/2}
e^{i  p^2 /4}  e^{-\pi  p^2/8}  D_{i p^2/2}
(-(1 - i) \tau)\,. \label{soltneg}
\end{eqnarray}

In the above expression, $A^-$ and $B^-$ denote the values of the
Bogolyubov coefficients before $\phi$ crosses $\phi_*$, while the
functions which multiply them are \emph{exact} solutions of the
\emph{linearized} equation~(\ref{eqapp}). The analytical approximation
consists in considering them as solutions of the true evolution
equation~(\ref{eom}). For $\tau \rightarrow + \infty$ it is
convenient to rewrite the solution~(\ref{soltneg}) in terms of two
different parabolic cylinder functions\footnote{This rewriting is
always possible since eq.~(\ref{eqapp}) has only two linearly
independent solutions.}~\cite{gr}
\begin{eqnarray}
&& u_+ ( \tau) = \left[ A^-
\frac{\sqrt{\pi}  p  e^{-\pi p^2/4}}{\Gamma \left( 1+ i p^2/2 \right)}
e^{i ( \frac{\pi}{4}-\frac{p^2}{2}+\frac{p^2}{2} \ln \frac{p^2}{2}
)} + B^-  e^{-\pi p^2/2} \right] \times \nonumber \\[5pt]
&&\hphantom{u_+ ( \tau ) =}
\times \left\{ \sqrt{2} \left( \frac{p}{\sqrt{2}} \right)^{-ip^2/2}
e^{ip^2/4}  e^{-\pi p^2/8}  D_{ip^2/2} ( ( 1 - i )\tau ) \right\} + \nonumber \\[5pt]
&&\hphantom{u_+ ( \tau ) =} + \left[ - A^- \, e^{-\,\pi\,p^2/2} + B^- \,
\frac{\sqrt{\pi} \, p \, e^{-\,\pi p^2/4}}{\Gamma
\left( 1- i\,p^2/2 \right)} \, e^{-\,i ( \frac{\pi}{4}-
\frac{p^2}{2}+\frac{p^2}{2} \ln \frac{p^2}{2} )}
\right] \times \nonumber \\[5pt]
&&\hphantom{u_+ ( \tau ) =} \times \left\{ \sqrt{2}
\left( \frac{p}{\sqrt{2}}
\right)^{1+\,i\,p^2/2}  e^{i( \frac{\pi}{4} -
\frac{p^2}{4} )} e^{-\pi p^2/8}  D_{-1-ip^2/2}
(( 1 + i) \tau ) \right\}.
\end{eqnarray}

In this new expression, the functions within curly brackets correspond to
the asymptotic forms at $\tau = +  \infty$ of the two terms of the
solution~(\ref{adiab}). The coefficients in front of them give thus the
new Bogolyubov coefficients in terms of the old ones.

All this derivation can be easily generalized when productions at
successive zeros of the total mass $m$ are considered. The only important
points are
\begin{list}{}{\setlength{\labelwidth}{20pt}}
\item[(i)] a difference in the values of the scale factor $a$ and of
the derivative $\phi'$ at different~$\eta_{*i}$'s,
\item[(ii)] a change of sign
in the transfer matrix (the one which gives the new coefficients in terms
of the old ones) whenever $\phi$ crosses $\phi_*$ from below to above
(cfr. the footnote just before eq.~(\ref{eqapp})), and
\item[(iii)] the phase
$e^{\pm  i \int \omega_k  d \eta}$ which accumulates between
$\eta_{*1}$ and the $\eta_{*i}$ considered.
\end{list}

Putting all this together, one has
\begin{eqnarray}
\pmatrix{\alpha_n \cr \beta_n^* } &=&
\pmatrix{F_n & H_n \cr -H_n^* & F_n^* }  \pmatrix{\alpha_{n-1} \cr
\beta_{n-1}^*} \quad \mathrm{\ for\ } n  \mathrm{\ odd} \,,
\nonumber\\[6pt]
H_n &\longleftrightarrow& -  H_n \hspace{3.8cm} \mathrm{\ for\ } n
\mathrm{\ even}\,, \label{matrix}
\end{eqnarray}
where $a_n , \beta_n$ are the values of the Bogolyubov
coefficients after the $n$-th production, and where
\begin{eqnarray}
F_n &=& \sqrt{1 - e^{-\pi p_n^2}}  e^{i( -  \frac{\pi}{4}
- \arg  \Gamma ( i  p_n^2 /2 ) + \frac{p_n^2}{2}
\ln{( p_n^2/2 )} - p_n^2/2 )}\,, \nonumber \\[10pt]
H_n &=& e^{- \pi p_n^2/2}  e^{- 2 i \int_{\eta_{*1}}^{\eta_{*n}}
\omega_k \, d \eta} \,,\qquad | F_n |^2 + | H_n |^2 = 1 \,. \label{defin}
\end{eqnarray}
We remind that $p_n = k / \sqrt{g  |
\phi_{*n}' | a_{*\,n}} $.

If one starts with no fermions at the beginning, one may choose
$\alpha_0 = 1, \beta_0 =~0$. Then, applying successive times the
``transfer'' matrix~(\ref{matrix}), one can get the spectrum of
fermions produced after every $\eta_{*n}$.

Of course our calculation reproduces the result
\begin{equation}
N_1= | \beta_1 |^2 = e^{-\pi p_1^2}
\end{equation}
reported in \pagebreak[3] ref.~\cite{ckrt}.

We numerically integrated the evolution equations for $u_\pm$,
$\phi$, and the scale \linebreak factor~$a$.\footnote{Our starting point
is at $\phi(0) = 0.28 M_p$, short after inflation,
$\phi' ( 0) = -0.15 M_p  m_\phi$ (as follows from a
numerical evaluation of the inflaton alone during inflation), and
$a ( 0 ) = 1$.} The results obtained with the
analytical expression~(\ref{matrix}) are always in very good
agreement with the numerical ones. Just to give a couple of
examples, we present here two cases at different regimes (we show
them only with illustrative purpose, and the values of the
parameters chosen have no particular importance). In
figure~\ref{spect1} we present the spectrum of the fermions after
two productions, that is after one complete oscillation of the
inflaton field. In analogy with the bosonic case, we measure the
strength of the coupling inflaton-fermions with the quantity $q
\equiv g^2  \phi_0^2 / \big( 4 m_\phi^2 \big)$,  where
$\phi_0 \simeq 0.28  M_p$ is the value of the inflaton at the
beginning of reheating. In figure~\ref{spect1} we choose $q=10^8$,
while we fix the bare fermion mass to be $m_X= 100  m_\phi$. In
figure~\ref{spect3} we show instead the resulting spectrum after
$7$ productions in the case $q=10^4$, $m_X=4  m_\phi $.

\FIGURE[t]{\epsfig{file=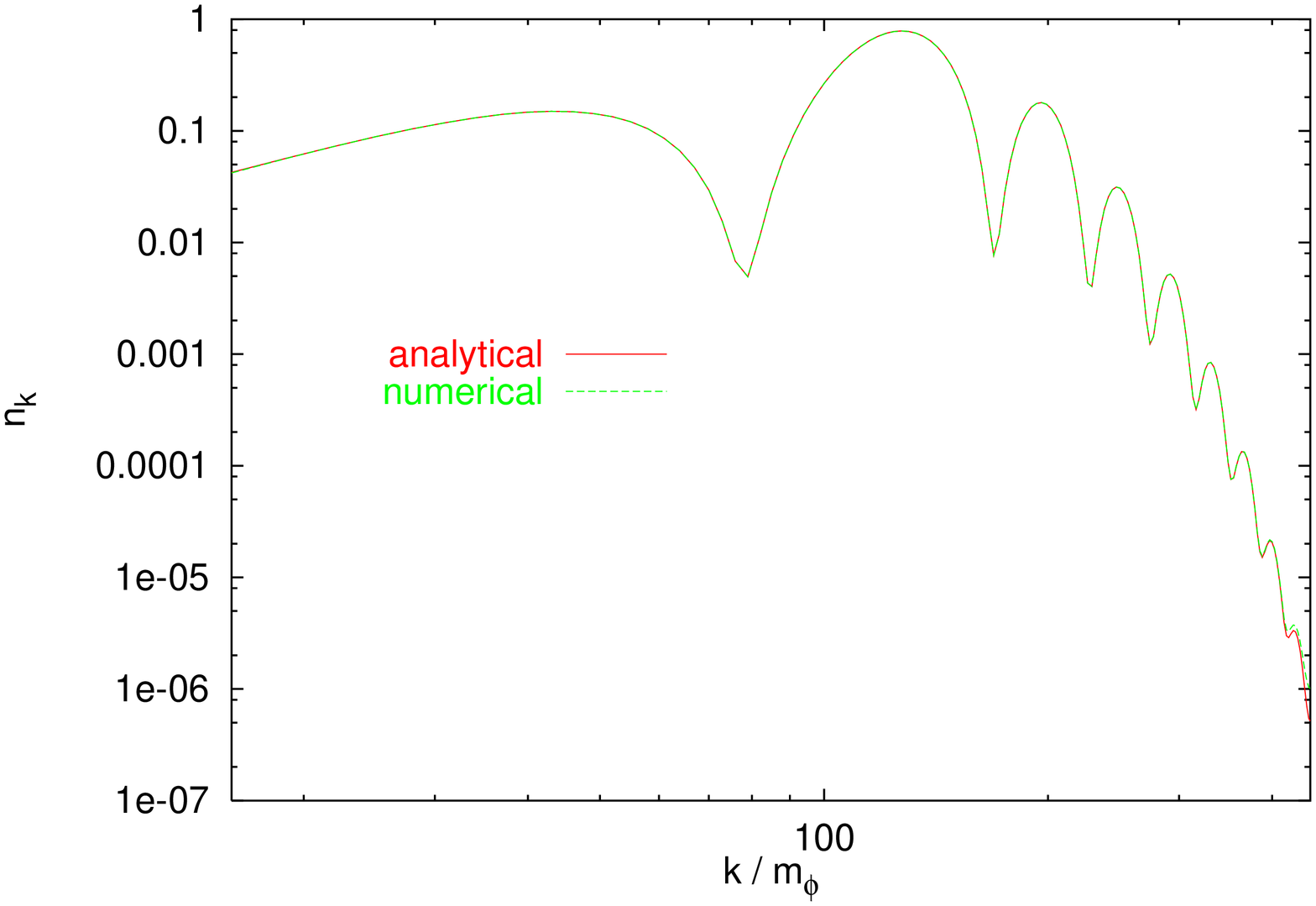, angle=-90, width=0.8\textwidth}
\caption{Spectrum of the fermions after two productions for $q=10^8$ and
$m_X=100m_\phi$. The expansion of the Universe is taken into
account.}\label{spect1}}

\FIGURE[t]{\epsfig{file=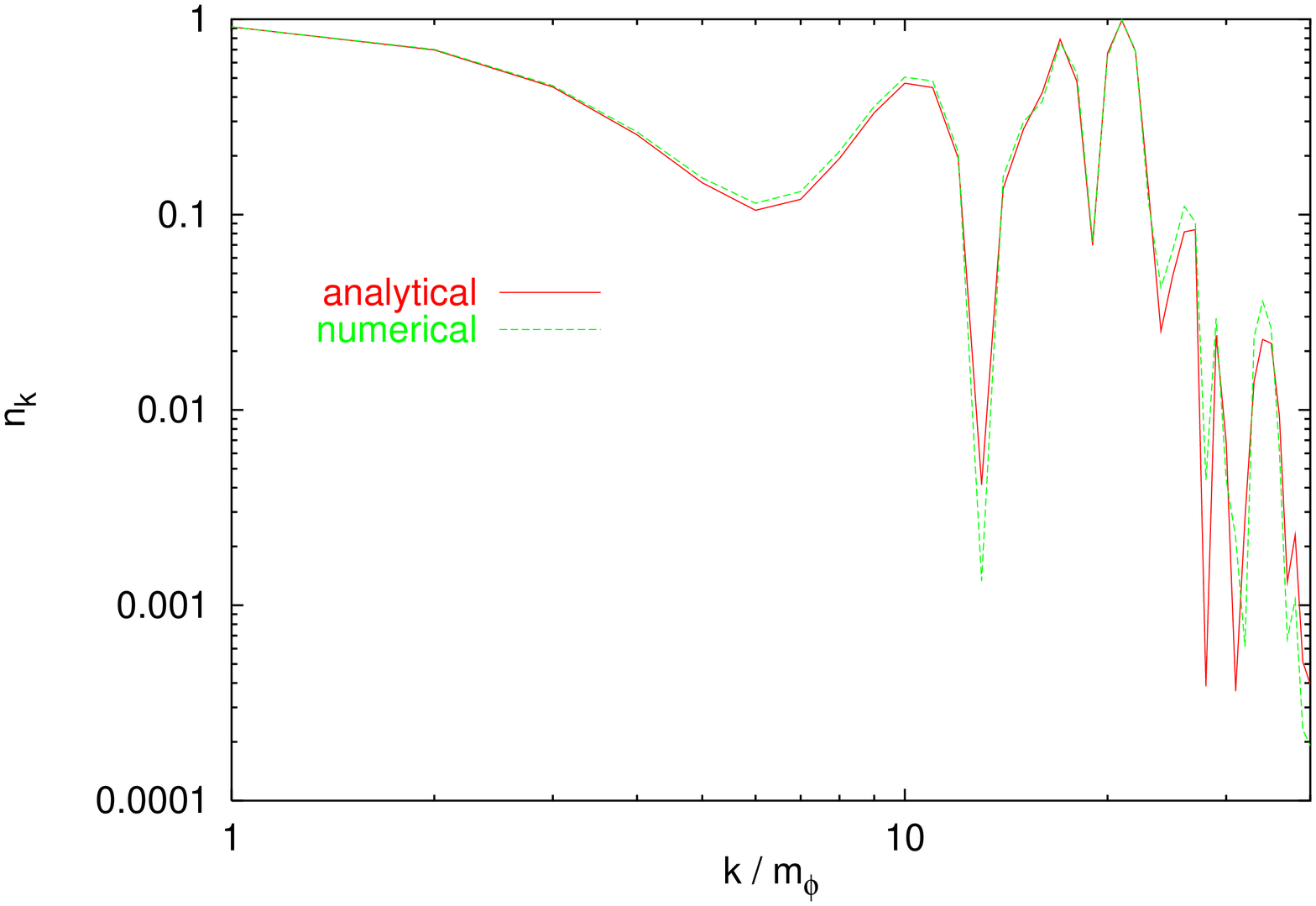, angle=-90, width=0.8\textwidth}
\caption{Spectrum of the fermions after seven productions for $q=10^4$ and
$m_X=4m_\phi$. The expansion of the Universe is taken into
account.}\label{spect3}}
\clearpage

\section{Production in a non-expanding Universe}\label{sec:4}
\vspace{-3pt}

In the bosonic case, the study of the non-perturbative production
in a non-expanding Universe has proven very useful in
understanding the effects of the production. It is shown in
ref.~\cite{kls} that the bosonic wave function satisfies the
Mathieu equations, whose solutions are characterized by resonance
bands (in momentum space) of very ``explosive'' and efficient
production. It is then shown that, due to the expansion of the
Universe, modes of a given comoving momentum $k$ cross several
resonance bands during the evolution. This gives the creation the
stochastic character described in the work~\cite{kls}. In
ref.~\cite{gk} it is understood that an analogous behavior occurs
also for fermions. The expansion is expected also in this case to
spoil the clear picture of distinct resonance bands. This fact may
help the transfer of energy to fermions, since the resonance bands
in the fermionic case are anyhow limited by the Pauli principle.
The expansion allows thus new modes to be occupied, and the
production is no longer limited to the regions of resonance. In
ref.~\cite{gk} it is stated that the production should then almost
completely fill the whole Fermi sphere up to a maximal momentum
$k_{\max}$. This behavior is confirmed by the numerical results of
the work~\cite{gprt}. In this section we will see that our
analytical formulae can reproduce the resonance bands, while in
the next one we will discuss the effects of the expansion of the
Universe.

Let us consider the matrices (we drop the index $n$ in the matrix
elements since all the $p_n$'s have now the same value $p$)
\begin{eqnarray}
M = \pmatrix{F & G \cr -  G & F^* } \,,\qquad
T_1 =\pmatrix{e^{i\,\vartheta_1^2} & 0 \cr 0 & e^{-i\vartheta_1^2}
}\,, \qquad T_2 = \pmatrix{e^{i\vartheta_2^3} & 0 \cr
0 & e^{- i \vartheta_2^3} } \,,
\end{eqnarray}
with $G \equiv e^{-\pi p^2/2}$ and $\vartheta_i^j \equiv
\int_{\eta_{*i}}^{\eta_{*j}} \omega_k  d \eta$ ($F$
was defined in the previous section).

Without the expansion of the Universe, the inflaton field has
the periodic \linebreak evolution
\begin{equation}
\phi ( \eta \equiv t ) = \phi_0  \cos{ \left( m_\phi  \eta \right)},
\end{equation}
and all the $\vartheta_i^j$~(\ref{matrix}) are hence sums of
$\vartheta_1^2$ and $\vartheta_2^3$ (remember the $\eta_{*i}$
are the moments at which the total fermionic mass vanishes).

After the generic $n$-th complete oscillation one thus has
\begin{equation} \label{op}
\pmatrix{\alpha_{2n} \cr \beta_{2n}^* } =
\pmatrix{ e^{-\,i\,\vartheta_1^{2 n+1}} & 0 \cr 0
& e^{i\vartheta_1^{2 n+1}} }  T_2 M^T T_1 M \cdots
T_2 M^T T_1 M  \pmatrix{1 \cr 0 }
\end{equation}
with the combination ${\hat O} \equiv T_2 M^T T_1 M$ repeated $n$
times.

One has thus simply to study the eigenvalue problem for ${\hat
O}$ (notice $\dett{\hat O}=1$). This operator has the form
\vspace*{-7pt}
\begin{eqnarray}
{\hat O} &=& \pmatrix{A & B \cr - B^* & A^*} \,,\nonumber\\
A &=&F^2  e^{i\left(\vartheta_1^2+\vartheta_2^3\right)}
+ G^2  e^{-i\left(\vartheta_1^2-\vartheta_2^3\right)}\,, \nonumber\\
B&=&F  G e^{i\left(\vartheta_1^2+\vartheta_2^3\right)}
- F^*  G  e^{-i\left(\vartheta_1^2-\vartheta_2^3\right)}
\label{op2}
\end{eqnarray}
and its eigenvalues are $\lambda_{1,2}=e^{\pm i \Lambda}$ with
$\cos \Lambda = {\Ree}\, A $.
\pagebreak[3]

Rewriting the initial condition {\scriptsize{\smash{$\pmatrix{1 \cr 0}$}}} in terms of
the eigenvectors of ${\hat O}$ and substituting in
formula~(\ref{op}), one gets the number of produced fermions
\begin{equation} \label{noexp}
N_n = | \beta_n |^2 =
\frac{ | B |^2}{1 - \left( \mathrm{Re\ } A \right)^2}
\sin^2 \left( n  \Lambda \right) =
\frac{ | B |^2}{\sin^2{\Lambda}}  \sin^2 \left( n  \Lambda \right)
\end{equation}
after the complete $n$-th oscillation.\footnote{In ref.~\cite{gk}
it is said that it is possible to extract the average over one
period of the occupation number from the knowledge of the solution
of eq.~(\ref{eom}) with initial conditions $u_+ \big(\eta_0
\big) = 1 , u_+' \big( \eta_0 \big) = 0$. The matching
of this procedure with our formulae~(\ref{adiab}) and~(\ref{op2})
gives the average time evolution (in our notation) $ {\bar N}
(t) = | B |^2 \, \sin^2 [( \pi -\Lambda) t / T ] / \sin^2
[ \pi - \Lambda]$, where $T$ is the period of one oscillation. Clearly,
this result coincides with our eq.~(\ref{noexp}) at any given
complete oscillation.}

We notice the presence of the envelope function
\begin{equation} \label{envelope}
{\tilde E} \equiv \frac{ | B |^2}{1 - \left( {\Ree}\, A
\right)^2} = \frac{ 1 - | A |^2}{1 - \left( {\Ree}\, A
\right)^2}
\end{equation}
which modulates the oscillating function $\sin^2 (n\Lambda)$.

With increasing $n$ this last function oscillates very rapidly and
can be at all effects averaged to $1/2$. One is thus left with
the envelope function which shows the presence of resonance bands.
The resonance bands occur where $A$ is real and ${\tilde E}
\rightarrow 1 $. It is easy to understand that their width
exponentially decreases with increasing momenta $k$. To see
this, let us consider the behavior of ${\tilde E}$ at high
momenta. In this regime, the function $A$ is given by
\begin{equation}
A \simeq ( 1 - e^{- \pi  p^2} )  e^{i  \phi_A} \,,
\qquad p =\frac{ k }{ \sqrt{g | \phi_*' |}} \,,
\end{equation}
where the phase $\phi_A$ can be read from eq.~(\ref{op2}). Near to
the points where $\cos \phi_A = 1 $ the envelope function
behaves like\footnote{A completely analogous behavior occurs where
$\cos \phi_A = -  1 $.}
\begin{equation} \label{width}
{\tilde E} \simeq \frac{1 - ( 1 - e^{ -  \pi p^2}
)^2}{1 - \left( 1 - e^{ -  \pi  p^2} \right)^2
\cos^2 \phi_A} \simeq \frac{1}{e^{\pi  p^2}
\left( 1 - \cos \phi_A \right) + 1} \,.
\end{equation}

The width of the band can be defined as the distance between the
two successive points at which ${\tilde E} = 1/2$. From the last
expression it follows that the difference between the phases of
$A$ in these two points is given by $\Delta \alpha = 2 \sqrt{2}
e^{-\pi p^2}$. Since the most rapidly varying term which
contributes to the phase of $A$ is $\big( p^2/2 \big) \log
\big( p^2/2 \big)$, the width of the band can be thus
estimated to be
\begin{equation}
\Delta p \simeq \frac{2 \sqrt{2}}{p  \log \left( p^2/2 \right)}\,
e^{-  \pi  p^2} \,.
\end{equation}

\FIGURE[t]{\epsfig{file=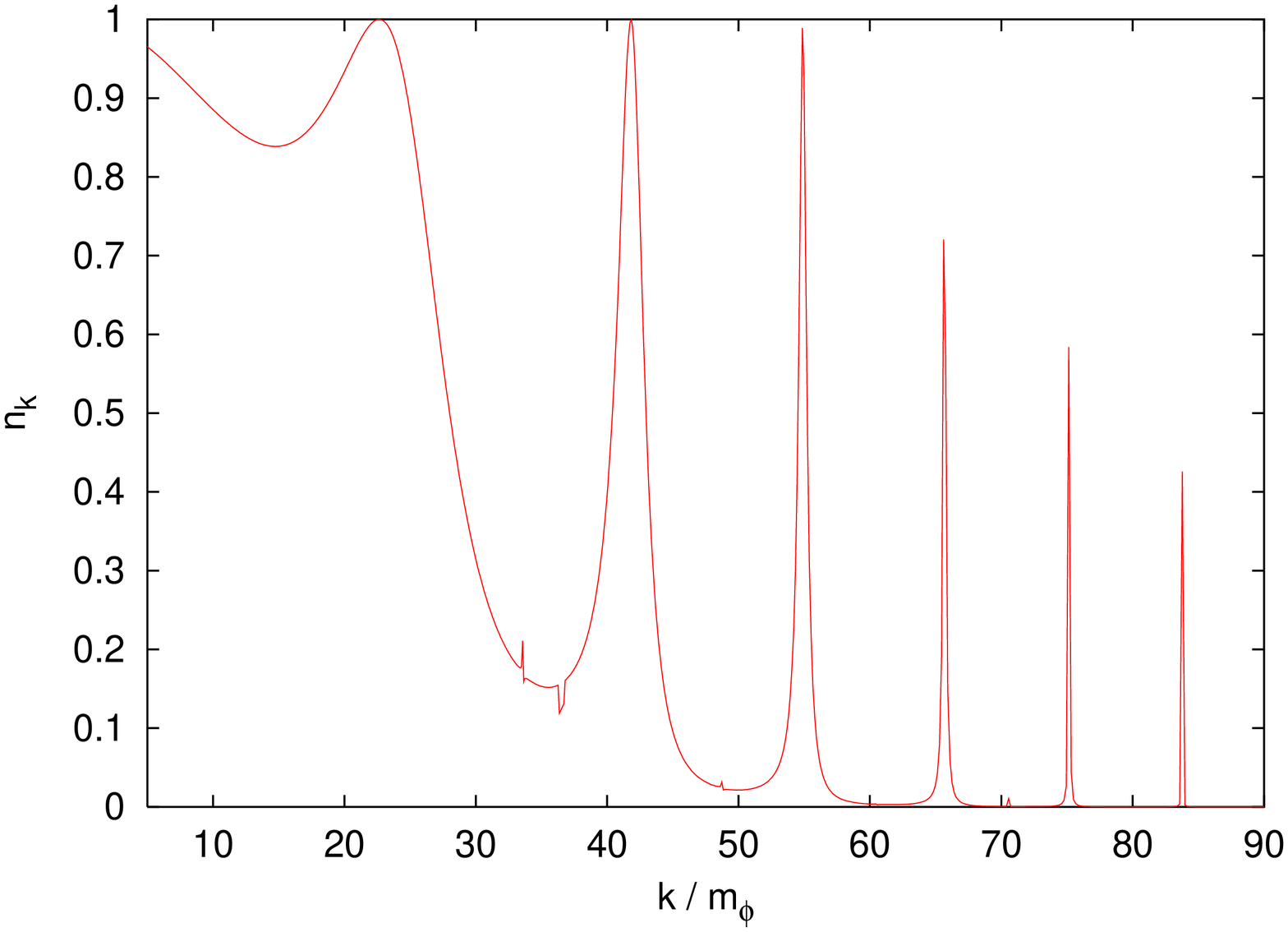, angle=-90, width=0.8\textwidth}
\caption{Envelope of the spectrum of the produced fermions in a static
Universe. The physical parameters are $q=10^6$ and $m_X = 100\,
m_\phi$.}
\label{static}}

We show in figure~\ref{static} the envelope of the produced
fermions in a static Universe for the parameters $q=10^6$ and $m_X
= 100\, m_\phi$. The peaks occur where $A$ is real and it is
confirmed that their width decreases very rapidly at increasing
momenta. Due to the fact that the last peaks plotted are indeed
very sharp, the resolution of figure~\ref{static} does not allow
to see their top at $n_k = 1$.

\section{Expansion taken into account}\label{sec:5}

As stated in the previous section, the resonance bands disappear
when the expansion of the Universe is taken into account. In this
section we will show how the occupation number varies when a
non-vanishing Hubble parameter is considered. As we have seen in
section~\ref{sec:2}, eqs.~(\ref{matrix}) and~(\ref{defin}) give a very
good agreement with the numerical results. On the other hand, the
presence of phases in eq.~(\ref{defin}) makes the exact analytical
treatment of the occupation number impossible. Now, the same
observation made in the bosonic case~\cite{kls} turns very useful
also to us: the phases in eq.~(\ref{defin}), when the expansion of
the Universe is taken into account, are not correlated among
themselves. As a result, the final spectra present several high
frequency oscillations about  some average function. The positions
of the peaks of these oscillations depend on the details of the
phases. However, the ``mean'' function can be understood in a
surprisingly easy way. Our problem can be treated as one customary
does when dealing with the ``random walk''. Imagine one has to
calculate the quantity
\begin{equation}
S \equiv | A_1 + A_2 + A_3 + \cdots + A_n |^2 \,,
\end{equation}
where the $A_i$ are complex numbers with random phases. The
``random walk recipe'' indicates that the best estimation of the
above quantity is achieved by summing the squares of the terms
$A_i$, since the mixed products average to zero. The chaoticity
of the final spectra for $n_k$ suggests that this may also be true
in our case, and this is confirmed by comparison with the
numerical results.

With this method, eqs.~(\ref{matrix}) and~(\ref{defin}) turn into
the much simpler relations
\begin{equation}
\pmatrix{| \alpha_n |^2\vspace*{5pt} \cr | \beta_n^*
|^2 }=\pmatrix{| F_n |^2 & | H_n |^2\vspace*{5pt} \cr
| H_n |^2 & | F_n |^2 }
\pmatrix{| \alpha_{n-1} |^2 \vspace*{5pt}\cr
| \beta_{n-1}^* |^2 }\,, \label{matrix1}
\end{equation}
where we remember
\begin{equation}
| F_n |^2 = 1 - e^{-\pi p_n^2}\,,\qquad
| H_n |^2 = e^{-\pi p_n^2}\,,\qquad
| F_n |^2 + | H_n |^2  = 1 \,. \label{defin2}
\end{equation}

Assuming no fermions in the initial state, and applying $n$ times
this formula, it is easy to see that the occupation number after
the $n$-th production is given by
\begin{equation}\label{prod}
N_n (k) =
\frac{1}{2}  -  \frac{1}{2} \prod_{i=1}^n
\left( 1 - 2  e^{- \pi  p_i^2} \right).
\end{equation}

A similar result holds for preheating of bosons, cfr.~\cite{kls}
where the idea of averaging on almost random phases is first
introduced. In the bosonic case, one can exploit the fact that,
due to the high efficiency of the production, the occupation
number after the $(n+1)$-th creation is (almost) proportional
to the occupation number after the $n$-th one:
\begin{equation}
N_{n+1} \left( k \right) \simeq
\left( 1 + 2  e^{\pi  \kappa_n^2} \right) N_{n} \left( k \right),
\end{equation}
where the quantity $\kappa_n$ is analogous to our parameter
$p_n$. This simplification is not possible in our case, since
the Pauli principle forbids $N_{n}$ to be sufficiently high.
However our final result, eq.~(\ref{prod}), is also cast in a very
simple and immediate form.

The validity of eq.~(\ref{prod}) is confirmed by our numerical
investigations, as we show here in one particular case. In
figure~\ref{guida} we compare the behavior of the ``mean''
function for the spectra with respect to the numerical one. We
choose the physical parameters to be $q=10^4 , m_X=4 m_\phi$, and
we look at the results after the $7$-th production (this
corresponds to the choices made in figure~\ref{spect3}). We see
that the ``mean'' function interpolates very well between the
maxima and the minima of the numerical spectrum, and that it can
indeed be considered as a very good approximation of the actual
result.

\FIGURE[t]{\epsfig{file=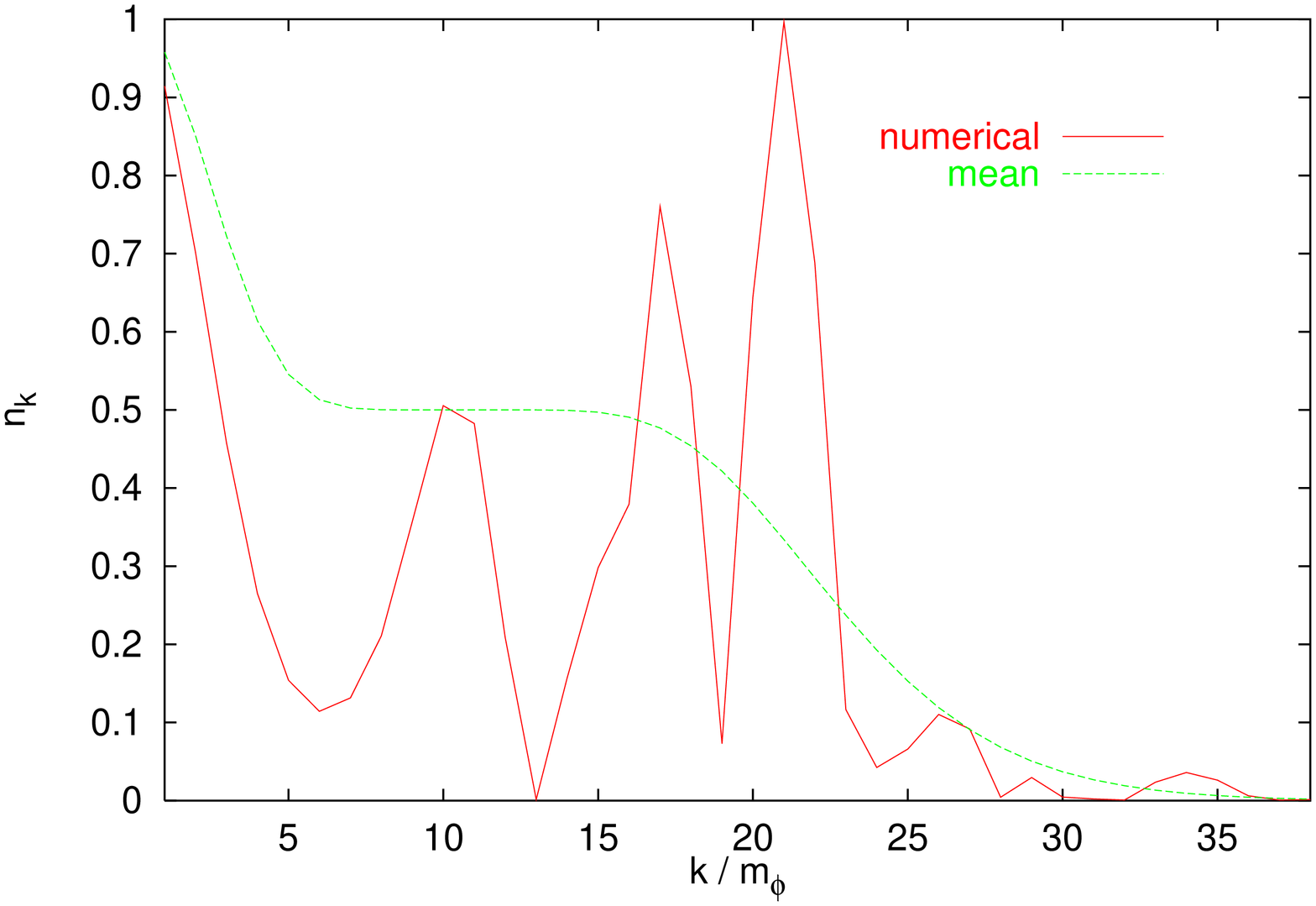, angle=-90, width=0.8\textwidth}
\caption{Comparison between the numerical spectrum and the ``mean''
function after seven productions for $q=10^4$ and $m_X=4
m_\phi$.}\label{guida}}

This is confirmed by figure~\ref{guida2}, where we plot (for the
same values adopted in figure~\ref{guida}) the quantity $n_k k^2$
rather than the occupation number alone. This quantity is of more
physical relevance when one is interested in the total energy
transferred to the fermions, since the total number density of
produced fermions is (apart from the dilution due to the expansion
of the Universe, that will be considered only in the final result)
\begin{equation}
N_X = \frac{2}{\pi^2}  \int d k\, k^2\,  n_k\,.
\end{equation}
As shown in the plot, the result for $N_X$ in the numerical and in
the approximated case are in very good agreement.

\FIGURE[t]{\epsfig{file=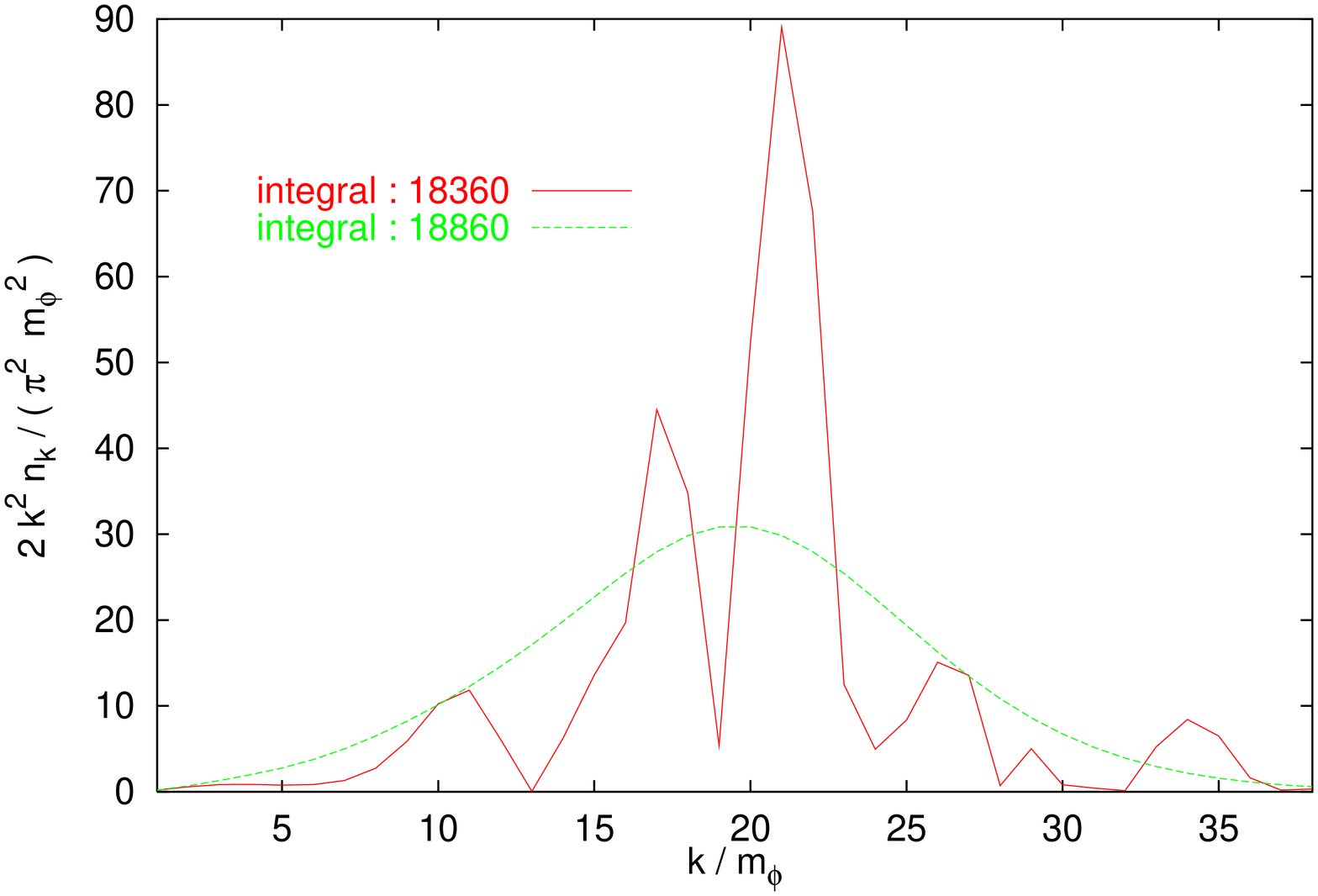, angle=-90, width=0.8\textwidth}
\caption{As in figure~\ref{guida}, but with the quantity
$k^2  n_k$ plotted.}\label{guida2}}

After checking the validity of the approximation given by the
``mean'' function, we adopt it to understand how the production
scales when different values of the parameters $q$ and $m_X$ are
considered.

Equation~(\ref{prod}) allows us to give an analytical estimate of the total
amount of energy stored in the fermions after the $n$-th production, and
in particular after that the whole process of non-perturbative production
has been completed. Notice that all the dependence on the physical
parameters is in the coefficients
\begin{equation}
z_i \equiv \frac{k}{ \left( \pi^{1/2}  p_i \right)}\,,
\end{equation}
where $k$ is the comoving momentum, and the only part we have to
determine are the numbers
\begin{equation} \label{zet}
z_i^2=\frac{2 \sqrt{q}}{\pi}a
\left(\eta_{*i}\right) \frac{\left|\phi'
\left(\eta_{*i}\right)\right|}{\phi_0}\,.
\end{equation}
Here, and in what follows, we express the dimensionful quantities
in units of $m_\phi$, the inflaton mass, apart from the inflaton
field $\phi$ that is given in units of $M_p$.

We also introduce the ratio $R \equiv 2 q^{1/2} / m_X $. This
quantity is the most relevant, since it determines the zeros of
the total fermionic mass and thus the values of the $z_i$'s.
Indeed, from eq.~(\ref{mass}) we see that the total mass vanishes
for $\phi_* = -  \phi_0 / R $. It is convenient to study the
production in terms of the two independent parameters $q$ and
$R$ (rather than $q$ and $m_X$) since, at fixed $R$, all the
spectra are the same provided we rescale $k \propto q^{1/4}$
(cfr. eq.~(\ref{zet})).

One can now proceed in two different ways, and we devote the next two
subsections to each of them. First, one can evolve the equations for the
inflaton field alone and find numerically the values $z_i$ for given $q$
and $R$. Inserting these  values in eq.~(\ref{prod}) one can get final
values for the production which, as we have reported, are very close to
the numerical ones. This method allows to get results which average the
actual ones, and it has the advantage of being much more rapid than a full
numerical~evolution.

Alternatively, one can proceed with a full analytical study in
order to understand the results given by the first semi-analytical
method.

\subsection{Semi-analytical results}

In this subsection we evaluate the analytical
formula~(\ref{prod}), taking the coefficients $z_i$ from a
numerical evolution of the inflaton field. As we have said, this
method gives results which are very close to the numerical ones,
due to the fact that the ``mean'' function interpolates very well
the actual spectra of the produced fermions.

The first thing worth noticing is that, for each choice of $q$ and
$R$, the maximum~$z_i$ occurs at about half of the whole process
of non-perturbative creation. It thus follows that fermions of
maximal comoving momenta will be mainly produced at the half of
the process. Our semi-analytical evaluations support this idea:
the total energy stored in the created fermions increases slowly
during the second part of the preheating. To see this, we show in
figure~\ref{intgrowth} the numerical results for the quantity $N_X
=2 \int d k\, k^2 n_k / \pi^2$ as function of the number of
\pagebreak[3]
production. We fix the physical parameters to be $q=10^6$ and
$R=30$, that is $m_X=67\,  m_\phi$. With this choice, the total
mass vanishes $10$ times, and figure~\ref{guida2} shows the results after
each step of this production. We see indeed that the final
productions are less efficient than the previous ones.

\FIGURE[t]{\epsfig{file=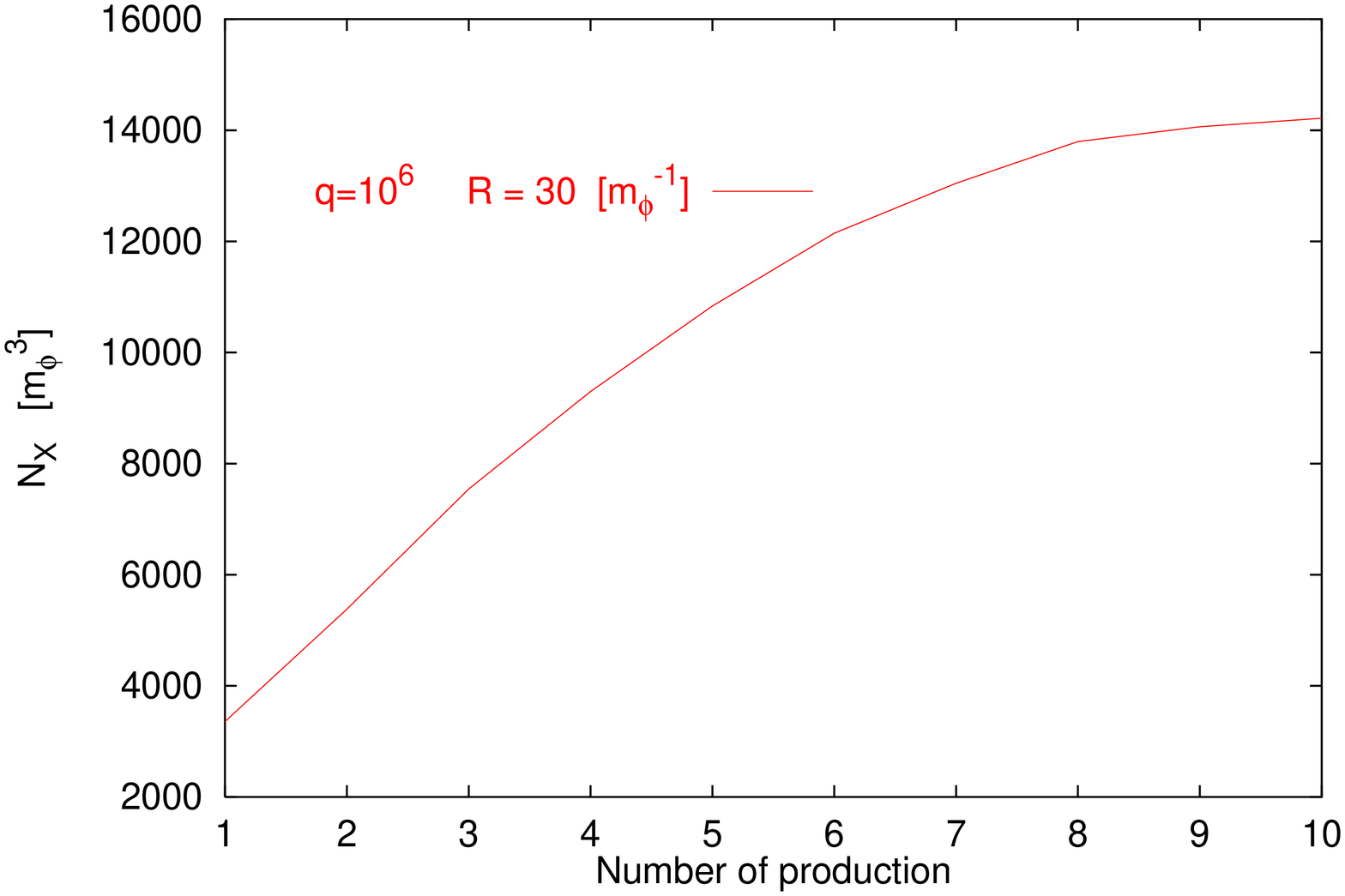, angle=-90, width=0.75\textwidth}
\caption{Growth of $N_X$ with the number of productions for $q=10^6$ and
$R=30$.}
\label{intgrowth}}

From eq.~(\ref{prod}) it is also possible to show the evolution of
the spectra with the number of productions. We do this in
figure~\ref{plotint}. We choose the same parameters as in
figure~\ref{intgrowth}, and we show the results after each
complete oscillation of the inflaton field (that is after each two
productions). We observe that the production rapidly approaches a
step function in the momentum space, i.e.\ there exists a maximum
momentum below which Pauli blocking is saturated (notice that the
value $1/2$ follows from the average understood in the ``mean''
function), and above which $n_k \simeq 0$. The fact that the
last productions do not contribute much to the total energy is
also confirmed.\footnote{The behavior at small $k$ is inessential,
since this region does not significantly contribute to the total
energy.}

{\renewcommand\belowcaptionskip{.6em}
\FIGURE[t]{\epsfig{file=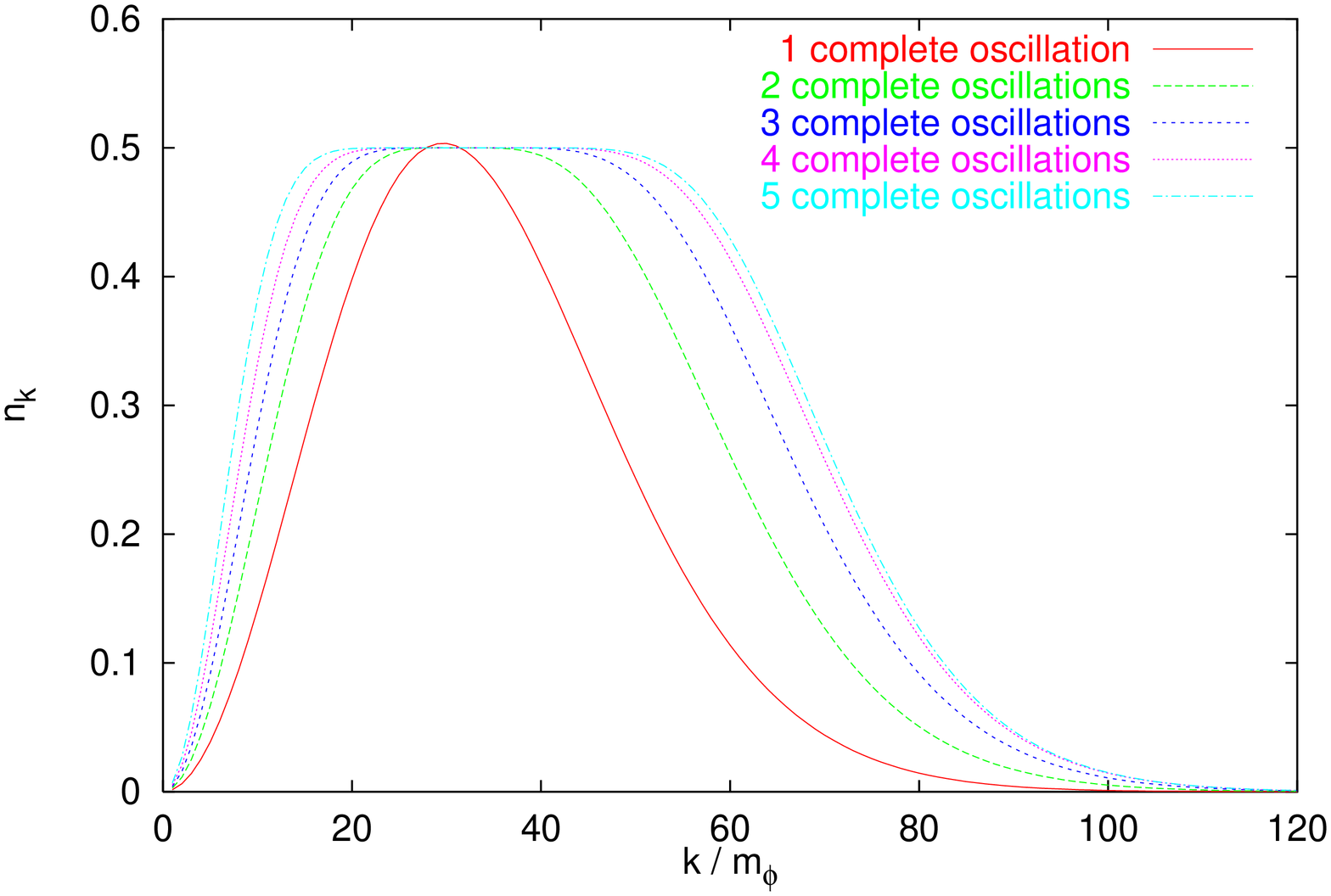, angle=-90, width=0.75\textwidth}
\caption{Evolution of the spectrum of produced fermions with the number of
productions for $q=10^6$ and $R=30$.}
\label{plotint}}

We now turn our attention to the total energy transferred to
fermions after the whole preheating process is completed. We fix
the parameter $q$ to the value $10^6$ and we investigate how the
total integral $N_X$ changes with different values for the
parameter $R$.\footnote{As reported, the scaling of the final
result with $q$ at fixed $R$ is simply understood from
eq.~(\ref{zet}).} The results are shown in figure~\ref{fitrho},
for $R$ ranging from $5$ to $10000$. For the last value the
total fermionic mass changes sign more than $3700$ times and a
full numerical evaluation would appear very problematic. This can
be done in our case, thanks to the analytical
expression~(\ref{prod}) found, and our results extend the validity
region of the previous full numerical study~\cite{gprt}.

\FIGURE[t]{\epsfig{file=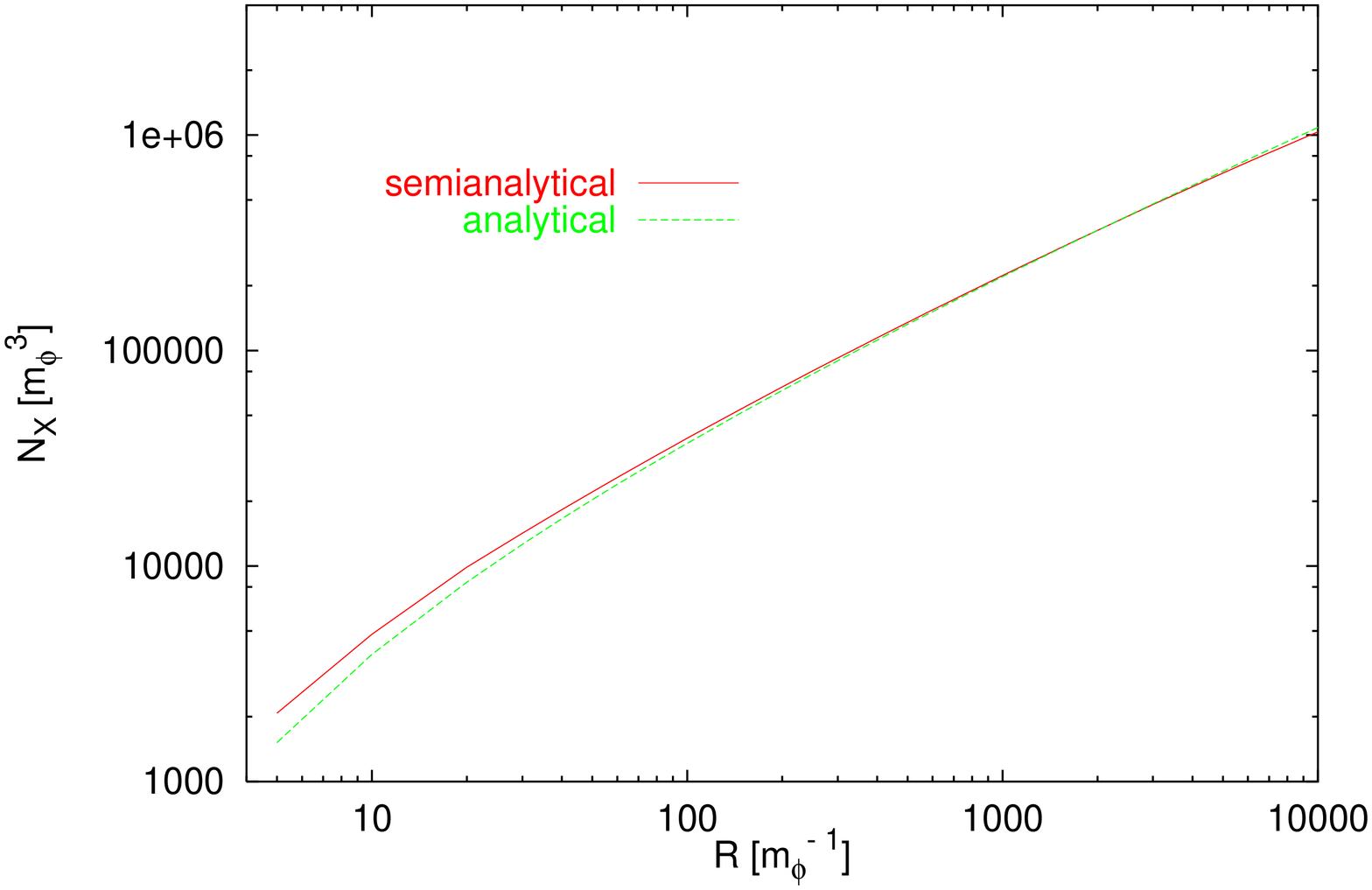, angle=-90, width=0.75\textwidth}
\caption{Comparison between the analytical and the semi-analytical results
for $N_X$, at fixed $q=10^6$.}
\label{fitrho}}}

In figure~\ref{fitrho}, the results of our semi-analytical method
are also compared to the full analytical ones of the next
subsection. This comparison will be \pagebreak[3] discussed below.

From the scaling of $N_X$ with $R$ just reported, it is now easy
to estimate the total energy transferred to fermions for generic
values of $q$ and $R$. We are interested in comparing our \pagebreak[3]
results to the numerical ones of the work~\cite{gprt}. To do so,
we consider the ratio between the energy density given to fermions
and the one in the inflaton~field\footnote{This ratio should be
calculated at a time $t_{\mathrm{end}}$ at the end of preheating,
when the total fermionic mass stops vanishing. Thus in the
denominator of eq.~(\ref{ratio}) the comoving momentum $k$ should
be replaced by the physical one $p=k/a$, with $a$ scale factor
of the Universe at $t_{\mathrm{end}}$. Analogously, the value
$\phi_0$ of the beginning of reheating should be replaced by the
one at $t_{\mathrm{end}}$. However, both the replacements cancel
out in the ratio, since both $\rho_\chi$ and $\rho_\phi$ redshift
as energy densities of matter.}
\begin{equation} \label{ratio}
\frac{\rho_X}{\rho_\phi} =
\frac{2  m_X}{\pi^2}  \int d k\,  k^2  n_k \cdot \frac{2}{m_\phi^2
\, \phi_0^2}
\,.
\end{equation}

We present our results in figure~\ref{rhotot}. We report them in
terms of $q$ and $m_X$, in order to have an immediate comparison
to the analogous figure~1 of ref.~\cite{gprt}.

\FIGURE[t]{\epsfig{file=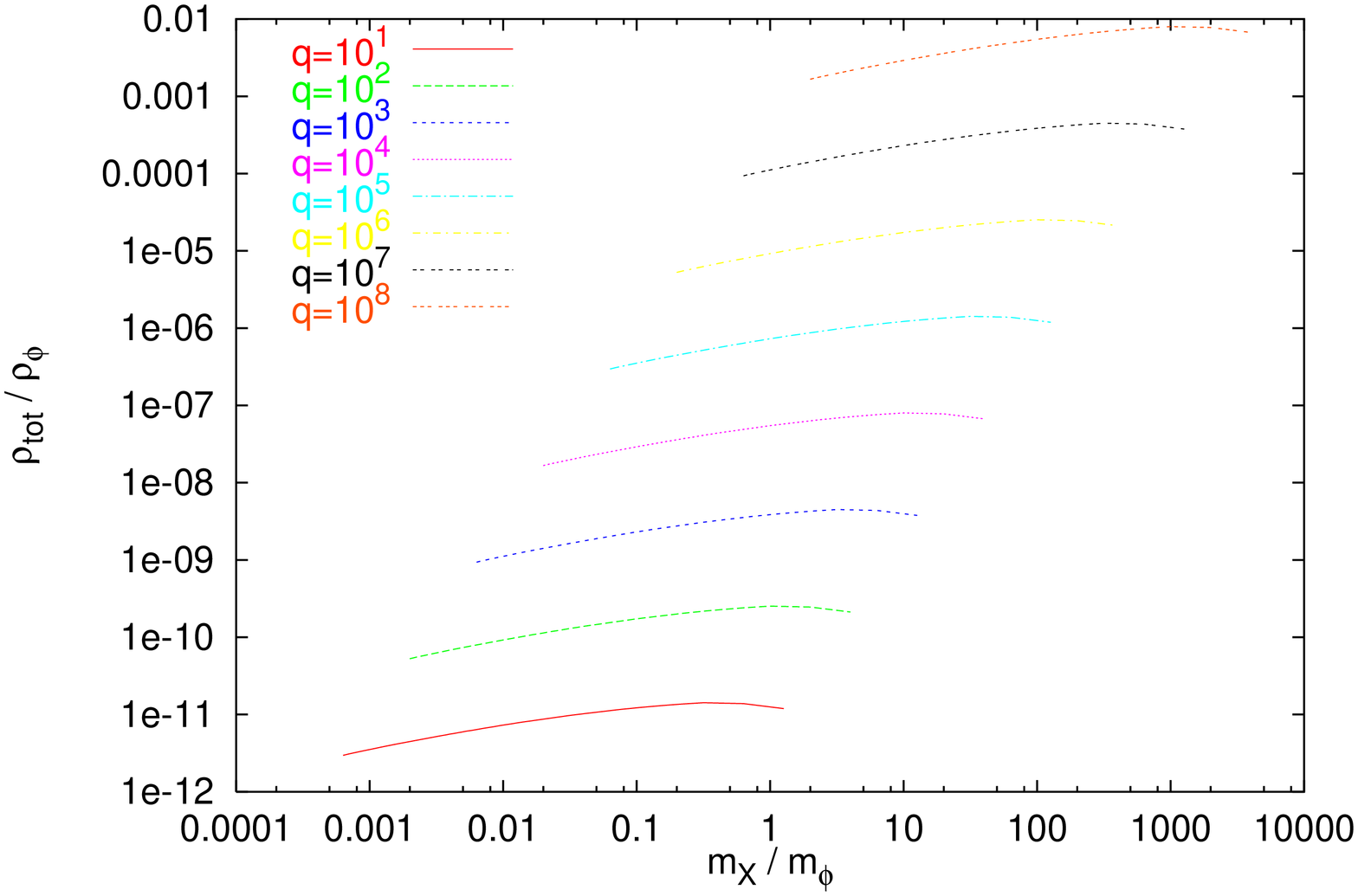, angle=-90, width=0.8\textwidth}
\caption{Total final energy density produced (normalized to the inflaton one)
for different values of $q$ and $m_X$.}
\label{rhotot}}

In figure~\ref{rhotot}, for any fixed $q$, the greatest plotted
value for $m_X$ corresponds to the choice $R=5$. We are not
interested in extending this limit since we know that for greater
$m_X$ (actually for values greater than the bound $m_X \sim
\sqrt{q}/2$) the production suddenly stops. The smallest value
plotted for $m_X$ (at any fixed value $q$) corresponds instead
to $R=10000$, that is to considering more than $3700$
productions in the numerical evaluation of eq.~(\ref{prod}).

Our final values are in good agreement with the ones of figure~1
of ref.~\cite{gprt} in the regime of validity of the latter. The
numerical results reported in that figure exhibit small
fluctuations about an average function $\rho_X \big(m_X\big)$. Our
results give this average function. This was expected, since the
expression that we integrated, eq.~(\ref{prod}), interpolates
between the maxima and the minima of the numerical spectra.

The numerical results of ref.~\cite{gprt} have a smaller range of
validity than the ones here presented. This occurs because the
numerical evolution of that work is limited to the first $20$
oscillations of the inflaton field, and so fermionic production
has not come to its end for small values of $m_X$. We confirm that
at high values of $m_X$ (actually at small values of $R$ for any
given $q$) the production depends very weakly on $m_X$. In
addition, our results show a decrease of the energy transferred to
fermions for smaller values of $m_X$. This behavior will be
explained in details in the next subsection.

\subsection{Analytical results}
We want now to show that all the results presented in the previous
section can be also achieved with a full analytical study of
eq.~(\ref{prod}).

First of all, we have to estimate the quantities $z_i$ given in
eq.~(\ref{zet}). To do this, it is more convenient to work in
terms of the physical time $t$: after the first few oscillations,
the inflaton evolution is very well approximated by the expression
(remember $t$ is expressed in units of $m_\phi^{-1}$, while
$\phi$ in units of $M_p$)
\begin{equation}\label{inflat}
\phi ( t) \simeq \frac{1}{\sqrt{3\pi}}
\frac{ \cos \left( t \right)}{t}\,.
\end{equation}
The scale factor of the Universe follows the ``matter-domination''
law, and it is well approximated by $a(t)=t^{2/3}$.

The values $t_{*i}$ are determined by the condition of vanishing
of the total mass of the fermions, that is, by making use of
eq.~(\ref{inflat}),
\begin{equation}\label{tstar}
\cos\left(t_{*i}\right)=-\frac{t_{*i}}{RA}\,,
\end{equation}
where we remind $R \equiv (2 \sqrt{q}) / m_X$. The parameter
$A\equiv(\sqrt{3 \pi}\, \phi_0)^{-1}$ is of order one and
will not play any special role in what follows. Notice that the
last production occurs at $t \simeq R A$.

Hence, keeping only the dominant contribution to the derivative of
$\phi$ with respect to the physical time, we get the expression
\begin{equation}\label{zetai}
z_i^2\cong\frac{2\sqrt{q}}{\pi}R^{1/3} A^{4/3}
\left(\frac{t_{*i}}{R\,A}
\right)^{1/3}\sqrt{1-\left(\frac{t_{*i}}{RA}\right)^2}\,,
\end{equation}
where we can assume $t_{*i}\simeq i \pi$.

Equation~(\ref{zetai}) exhibits a very good agreement with the numerical
evaluation of the same quantity. It also shows that the maximal value for
$z_i$ is reached at $t_{*i}=A R/2$, that is, at half of the whole
process of non-perturbative creation. This was anticipated in the previous
section, where we showed that the most of the fermionic production occurs
in the first part of preheating.

Starting from eq.~(\ref{zetai}) we can also calculate the number
density of produced particles
\begin{equation}
N_X ( q , R ) =\frac{2}{\pi^2}  \int d k \, k^2  N (k ) \,,
\end{equation}
where $N(k)$ is obtained from eq.~(\ref{prod}) with \pagebreak[3]
$n=n_{\max}=(RA)/\pi$.

For $R$ big enough, the product in eq.~(\ref{prod}) can be written
as the exponential of an integral. Thus, we obtain
\begin{eqnarray}
N_X &=& \frac{m_X}{4  \pi^3}
\left( \frac{2 A^{4/3}}{\pi} \right)^{3/2}
q^{3/4}  R^{1/2} \times \label{integrali}\\ &&{}\times \int_0^\infty dt\,  t^2
\left\{ 1
- \exp \left[ \frac{R\,A}{\pi} \int_0^1 dy  \log
\left| 1 - 2 \exp \left( - \frac{t^2}{y^{1/3} \sqrt{1-y^2}}
\right) \right|\; \right] \right\}, \nonumber
\end{eqnarray}
with the substitution $t= k \cdot \sqrt{\pi / \big( 2  A^{4/3}
 q^{1/2}  R^{1/3} \big)}$.

The integral in $dy$ which appears in eq.~(\ref{integrali}) cannot
be calculated analytically. Anyhow, we can approximate it by
\begin{equation}
\int_0^1 dy \log \left| 1 - 2 \exp
\left( - \frac{t^2}{y^{1/3} \sqrt{1-y^2}} \right) \right|
\simeq g \left( t \right)  \log \left| 1 - 2  e^{-1.5  t^2}
\right|,
\end{equation}
where $g(t)$ is a function of order one that, for our
purposes, can be approximated by a constant $c$ in the range
$0.5\lesssim  c\lesssim 1$.

Hence, the integrand within curly brackets in
eq.~(\ref{integrali}) rewrites
\begin{equation} \label{occu}
1 - \left|  1-2 e^{-  1.5  t^2}  \right|^{c \frac{R A}{\pi}}.
\end{equation}
In the large-$R$ limit this function approximates a step function,
which evaluates to one for
\begin{equation} \label{kappa}
\sqrt{\frac{\pi \log 2}{2 R A c}}\lesssim  t
\lesssim \sqrt{\log\left(\frac{2 R A c}{\pi \log 2}\right)}
\end{equation}
and to zero for the remaining values of $t$.

Since the quantity~(\ref{occu}) is proportional to the occupation
number $n_k$, our analytical calculation confirms the usual
assumption that, after few oscillations, the fermionic production
saturates the Fermi sphere up to a given maximum momentum
$k_{\max}$. This is also shown in the previous
figure~\ref{plotint}. However, our derivation gives a different
scaling for $k_{\max}$ with respect to the previous
literature~\cite{gk,gprt}. Indeed, from eq.~(\ref{kappa}) it
follows (apart from proportionality factors)
\begin{equation} \label{kmax}
k_{\max} \propto \frac{q^{1/3}}{m_X^{1/6}}\,
\sqrt{ \log \left( \frac{q^{1/2}}{m_X} \right)},
\end{equation}
which is however quite close to the result given in~\cite{gprt}.
We will comment more about the origin of the scaling~(\ref{kmax})
in the conclusions.

From eqs.~(\ref{integrali}) and~(\ref{kappa}), we obtain the final
expression for the number density of the fermions created during the
whole process
\begin{equation} \label{fit}
N_X \left( q, R \right) = \frac{1}{3  \pi^2}
\left(\frac{2  A^{4/3}}{1.5  \pi}\right)^{3/2}q^{3/4}  R^{1/2}
\left[\log \left( \frac{4  A  c}{\pi  \log 2}  \frac{q^{1/2}}{m_X}
\right)\right]^{3/2}.
\end{equation}

We can now go back to figure~\ref{fitrho}, where this last
equation (called ``analytical'' in the figure) is compared to the
result with the semi-analytical method of the previous subsection.
In plotting eq.~(\ref{fit}) we chose $c=0.78$ for the numerical
factor involved. As we can see, the final results achieved with
the two methods are in very good agreement with each other, thus
confirming the validity of the formula~(\ref{fit}).\footnote{The
small discrepancy between the two curves can be attributed to the
fact that $c$ is not exactly constant.}

Rewriting eq.~(\ref{fit}) in terms of $q$ and $m_X$ we can draw
some conclusions. First, apart from a logarithmic correction, the
scaling of the total energy
\begin{equation}
\rho_X \propto m_X  N_X \propto q  m_X^{1/2}
\left[\log \left( \frac{4  A  c}{\pi  \log 2}\,
\frac{q^{1/2}}{m_X}  \right) \right]^{3/2}
\end{equation}
is linear in $q$, as generally expected~\cite{gprt}. The
dependence of $\rho_X$ on $m_X$ requires some more care: the
threshold value for $m_X$ is given by the condition $R \sim 4$,
that is,\footnote{The number $4$ comes from the fact that the
value of the inflaton at its first minimum is $\phi \simeq - 0.07
 M_p$, while at beginning $\phi_0 = 0.28  M_p$.}
\begin{equation} \label{thr}
\left(m_X\right)_{\mathrm{th}}\sim \frac{\sqrt{q}}{2} \,.
\end{equation}
For values of $m_X$ not much smaller than $(m_X)_{\mathrm{th}}$,
the total energy depends very weakly on $m_X$, this result being
in agreement with the numerical evaluation given in~\cite{gprt}.
On the other hand, for values of $m_X$ much smaller than
$(m_X)_{\mathrm{th}}$, the factor $m_X^{1/2}$ starts to dominate,
and we expect it to determine the scaling of the total energy when
$m_X\rightarrow 0$.

\section{Backreaction}\label{sec:6}

The results presented so far have been achieved neglecting the
backreaction of the produced fermions on the evolution of the
inflaton field and on the scale factor. This is a common
approximation, since a more complete treatment (especially an
analytical one) of the whole phenomenon is a very difficult task.
However, the effects of the backreaction can be understood to a
good degree of accuracy in the Hartree approximation.

For what concerns preheating of fermions, the Hartree
approximation consists in taking into account the term
\begin{equation} \label{hart}
g  \langle {\bar X}  X \rangle
\end{equation}
into the evolution equations for the inflaton and the scale
factor. The equation for the field $\phi$ thus rewrites (in
physical time)
\begin{equation} \label{binf}
\ddot{\phi} + 3  H  \dot{\phi} + m_\phi^2
 \phi + g  \langle {\bar X}  X \rangle = 0 \,.
\end{equation}

The study of this effect has been performed numerically in
ref.~\cite{gprt}, where it is shown that backreaction starts to be
important for  $q > q_b \sim 10^8 - 10^{10}$. Figure~7
of that work shows how the evolution of the inflaton field $\phi$
is modified when backreaction is considered and $q$ is
sufficiently high. First, one observes that the amplitude of the
oscillations of the inflaton is very damped already after the
first production. This effect is the most obvious one, since the
term~(\ref{hart}) takes into account the decay of the inflaton
into fermion-antifermion pairs, while in its absence the equation
for $\phi$ considers only the damp due to the expansion of the
Universe. The second feature that emerges from the evolution
performed in ref.~\cite{gprt} is that at the beginning the
field~$\phi$ does not oscillate about the minimum of the potential $V =
m_\phi^2 \phi^2 /2$, but about the point $\phi_*$ where the
total fermionic mass vanishes. Moreover, the frequency of these
oscillations is higher than $m_\phi $.

These last two effects are due to the change in the effective
potential for $\phi$ induced by the term~(\ref{hart}), and
disappear when the quantity $\langle {\bar X} X \rangle$ is
decreased by the expansion of the Universe. Their net effect is to
render the whole mechanism of preheating more efficient, since the
rise in the frequency of the oscillations of $\phi$ increases the
number of productions of fermions. In ref.~\cite{gprt} it is
indeed shown that for $q = 10^8$ the total production is about $5
\%$ larger than the one without backreaction, while for
$q=10^{10}$ the increase is about $50 \%$.

We now briefly study the evolution of the inflaton $\phi$ under
eq.~(\ref{binf}) by means of the analytical results presented
above. We show that even a very approximate analysis confirms the
numerical result that indicates in $q \simeq 10^8 - 10^{10}$ the
threshold above which backreaction should be considered.

To begin on, the term~(\ref{hart}) needs to be normal ordered.
Doing so, one gets~\cite{gprt}
\begin{equation}
\langle {\bar X}  X \rangle = \frac{2}{\left(
2  \pi  a \right)^3} \int d^3 {\bf k} \left( 1 + \frac{m
a}{\omega} - | u_- |^2 \right).
\end{equation}
In terms of the Bogolyubov coefficients, this quantity evaluates
to
\begin{equation} \label{back}
\langle {\bar X}  X \rangle =
\frac{4}{\left( 2  \pi  a \right)^3}
\int d^3 {\bf k} \left[ | \beta |^2
\frac{a  m}{\omega} + \Ree \left( \alpha  \beta \frac{k}{\omega}
\,e^{2  i  \int \omega \, d \eta} \right) \right].
\end{equation}

We see that $\langle {\bar X} X \rangle$ vanishes for $\beta =
0$. This is obvious, since backreaction starts only after
fermions are produced. Some approximations can render
eq.~(\ref{back}) more manageable. First, we notice that the
oscillating term in the exponential averages to very small values
the integral of the second term in square brackets. Second, we see
that $k \ll | a  m |$ where $\beta_k$ is significantly
different from zero. From both these considerations, the integrand
in eq.~(\ref{back}) can be approximated (up to the sign of~$m$)
by the occupation number $| \beta |^2$, so that the
whole effect is (approximatively) proportional to the number of
produced fermions.

The numerical results of ref.~\cite{gprt} show that, for the
values of $q$ for which the backreaction is to be considered, its
effects can be seen already in the first oscillation of the
inflaton field. Since we are only interested in estimating the
order of magnitude of $q_b$, we thus concentrate on the first
oscillation of $\phi$, neglecting the expansion of the Universe
in this short interval.

\looseness=-1
With all these approximations, eq.~(\ref{binf}) rewrites
\begin{equation} \label{beq}
y'' + y + 10^{-12}  q^{5/4} \frac{m}{ | m |} = 0 \,,
\end{equation}
where we have rescaled $y \equiv \phi / \phi_0$ and we remind that
the time is given in units of~$m_\phi^{-1}$.

This equation is very similar to the one obtained in the bosonic
case~\cite{kls}. The last term changes sign each time $m=0$ and,
when sufficiently high, forces the inflaton field to oscillate
about the point $\phi_*$ at which the total fermionic mass
vanishes. It is also responsible for the increase of the frequency
of the oscillations. To see this, we assume that this term
dominates over the second one of eq.~(\ref{beq}) and we solve it
right after the first fermionic production at the time $t_{*1}$.
In the absence of the second term, eq.~(\ref{beq}) is obviously
solved by a segment of parabola until the time $t_{*2}
\simeq t_{*1} + 2 | y' \big( t_{*1} \big) | /
\big(q^{5/4}  10^{-12} \big)$ at which $m$ vanishes again
and the last term of eq.~(\ref{beq}) changes sign. As long as the
second term of eq.~(\ref{beq}) can be neglected, the inflaton
evolution proceeds along segments of parabola among successive
zeros of the mass $m$. The ``time'' duration of these segments is
expected to be of the same order of the first one, since the
successive fermionic productions balance the decrease of $\langle
{\bar X}  X \rangle$ due to the expansion of the
Universe.\footnote{However, after the first part of the process,
the production looses its efficiency and the expansion of the
Universe dominates. As we have said, the term $\langle {\bar X} X
\rangle$ can then be neglected and the inflaton starts oscillating
about the minimum of the tree level potential.}

The period of these oscillations can thus be roughly estimated to
be $T \sim \linebreak 2 \big( t_{*2} - t_{*1} \big)$. We see that, for $q
\gtrsim 10^9 $, this period is smaller than the one that the
inflaton oscillations would have neglecting backreaction. Since
this increase of the frequency is the main responsible for the
higher fermionic production, the result $q_b \sim 10^9$ can be
considered our estimate for the value of $q$ above which
backreaction should be taken into account.

This result, although obtained with several approximations, is in
agreement with the numerical one of ref.~\cite{gprt}.

\section{Conclusions}\label{sec:7}

Preheating of fermions and bosons present several analogies. In
both cases, especially when one is interested in the production of
very massive particles, the creation occurs for very short
intervals of time, during which the frequency of the particles
varies non adiabatically and their occupation number cannot be
defined. This strong similarity can be exploited to extend to the
fermionic case the formalism developed in ref.~\cite{kls} for the
analytical study of preheating of bosons. This was first done in
ref.~\cite{ckrt}, where the result for the particles created in
the first production is reported. \pagebreak[3] When one wants to consider the
successive productions, a more detailed analysis is necessary.
This has been the object of the first part of the present work.
The formulae that we obtained, valid for an arbitrary number of
productions, give a very good agreement with the numerical
results, as some examples provided manifest.

Another strong similarity between the two phenomena is that in
both cases the production would occur only through resonance bands
(in momentum space) was not for the expansion of the Universe. For
what concerns fermions, this feature was first studied in
ref.~\cite{bhp,gk}, where production in a static Universe is
considered. As it is well known, the expansion of the Universe
removes the resonance bands and, as a result, this allows the
fermionic production to saturate the whole Fermi sphere up to a
maximal momentum $k_{\max}$. Our analytical results confirm both
the presence of the resonance bands in the static case (where our
formulae considerably simplify) and their disappearance when the
expansion is considered.

Rather than the detailed form of the spectrum, a quantity which may be
of more physical interest is the total energy of the produced
particles. The analogy with the bosonic case turns useful also in this
regard. In the final spectra one notices the presence of some maxima
and minima, whose exact position is determined by the detailed
knowledge of the phases of the Bogolyubov coefficients (adopted in the
analytical derivation) after each production. However, due to the
expansion of the Universe, these phases are effectively uncorrelated
among themselves. This gives the production a stochastic character
which is responsible for the disappearance of the resonance
bands. Averaging over these phases, it is possible to get an
analytical function for the mean occupation number. For what concerns
preheating of bosons, the results are simplified by the fact that, in
the region of very efficient production, the mean occupation number
after a given production is proportional to the mean occupation number
of the previous one. This assumption is not possible in the fermionic
case, since the Pauli principle prevents the occupation number to
exceed $1$. However, it is remarkable that also for fermions the final
result can be cast in a very simple and readable form.

The analytical formula for the mean occupation number confirms the
saturation of the Fermi sphere (times the factor $1/2$ which comes
from the average) up to a momentum $k_{\max}$. In order to get
this quantity, it is sufficient to know the derivative of the
inflaton field and the value of the scale factor at the points
where the productions occur. These values can be calculated
numerically evolving the equations for the inflaton field alone
(at least when backreaction is neglected), or analytically. The
former possibility is much more rapid than a full numerical
computation, since one does not have to consider the equations for
the fermions. Moreover, its results agree very well with the full
numerical ones~\cite{gprt}. Thanks to the increase in the
rapidity, this evaluation (that is called ``semi-analytical'' in
the present work) allows to get results with a more extended range
of validity with respect to the previous full numerical~ones.

The results provided in this way can also be achieved with a full
analytical study. From both these methods, we deduce that the
ratio $\rho_X / \rho_\phi$ scales as
\begin{equation} \label{final}
\frac{\rho_X}{\rho_\phi} \sim q m_X^{1/2} \left[ \log  \frac{q^{1/2}}{m_X}
\right]^{3/2},
\end{equation}
up to $\big( m_X \big)_{\mathrm{th}}\sim \sqrt{q}/2$.

For $m_X$ not too smaller than $\big( m_X \big)_{\mathrm{th}}$
this scaling is in good agreement with the one $\rho_X / \rho_\phi
\sim q$ of the numerical work~\cite{gprt}. However, we see that
the density of produced fermions decreases at smaller values of
$m_X$.

We would like to conclude our work showing that the
scaling~(\ref{final}) can be also achieved from very immediate
considerations. As we said, preheating of fermions saturates a
Fermi sphere in momentum space up to a maximal momentum
$k_{\max}$. From the analytical formula~(\ref{matrix1}), we
notice that at high momenta $k$ the occupation number is well
approximated by
\begin{equation}
N_n ( k ) \simeq \sum_{i=1}^n e^{- k^2/z_i^2} \,,
\end{equation}
where we remember $z_i \propto q^{1/4}  a^{1/2}
\big(\eta_{*i}\big)  | \phi'\big(\eta_{*i}\big)
|^{1/2}$.

\looseness=-1
In this last equation, we replace all the parameters $z_i$ with a
mean value ${\bar z}$, so that $N_n \sim n\,  \mathrm{exp}
\big( - k^2 / {\bar z}^2 \big)$. The scaling of ${\bar
z}$ with the physical parameters $q$ and $m_X$ follows from the
scaling of all the $z_i$. The maximal momentum $k_{\max}$ is thus
expected to scale as the quantity $z_i  ( \log n)^{1/2}$.
Considering now the evolution of the inflaton field in physical
time $t$, we notice that both the number $n$ of productions and
the times $t_{*i}$ at which they occur are proportional to the
parameter $R = q^{1/2} / \big( 2 m_X)$. Moreover, we see that the
$z_i$'s scale as
\begin{equation}
z_i \propto  q^{1/4}  a_{*i}  \left[ \frac{d \phi}{ d  t}
|_{*i}  \right]^{1/2} \propto q^{1/4}
t_{*i}^{2/3}  \frac{1}{t_{*i}^{1/2}} \propto q^{1/4}
R^{1/6} \,.
\end{equation}

We thus get $k_{\max} \propto q^{1/4}  R^{1/6} [ \log R]^{1/2}$,
from which (remember $\rho_X \simeq m_X  k_{\max}^3$) the
scaling~(\ref{final}) simply follows.

\acknowledgments

We would like to thank Antonio Riotto for some suggestions and for
very useful discussions.

\end{document}